\pdfoutput=1
\documentclass[12pt,a4paper,pdftex]{article}

\usepackage{ifthen}    % for conditional statements
\newboolean{pdflatex}
\setboolean{pdflatex}{true} % False for eps figures

\newboolean{articletitles}
\setboolean{articletitles}{true} % False removes titles in references

\newboolean{uprightparticles}
\setboolean{uprightparticles}{false} %Set to true to get roman particle symbols

\usepackage{microtype}
\usepackage{lineno}  % for line numbering during review
\usepackage{xspace} % To avoid problems with missing or double spaces after
\usepackage{graphicx}  % to include figures (can also use other packages)
\usepackage{color}
\usepackage{colortbl}
\graphicspath{{./figs/}} % Make Latex search fig subdir for figures

\usepackage{tabularx}
\usepackage{dcolumn}
\newcolumntype{d}[1]{D{.}{.}{#1}}
\newcolumntype{C}{>{\centering}X}
\usepackage{calc}

\usepackage{amsmath} % Adds a large collection of math symbols
\usepackage{amssymb}
\usepackage{amsfonts}
\usepackage{upgreek} % Adds in support for greek letters in roman typeset

\newcommand*\patchAmsMathEnvironmentForLineno[1]{%
\expandafter\let\csname old#1\expandafter\endcsname\csname #1\endcsname
\expandafter\let\csname oldend#1\expandafter\endcsname\csname
end#1\endcsname
 \renewenvironment{#1}%
   {\linenomath\csname old#1\endcsname}%
   {\csname oldend#1\endcsname\endlinenomath}%
}
\newcommand*\patchBothAmsMathEnvironmentsForLineno[1]{%
  \patchAmsMathEnvironmentForLineno{#1}%
  \patchAmsMathEnvironmentForLineno{#1*}%
}
\AtBeginDocument{%
\patchBothAmsMathEnvironmentsForLineno{equation}%
\patchBothAmsMathEnvironmentsForLineno{align}%
\patchBothAmsMathEnvironmentsForLineno{flalign}%
\patchBothAmsMathEnvironmentsForLineno{alignat}%
\patchBothAmsMathEnvironmentsForLineno{gather}%
\patchBothAmsMathEnvironmentsForLineno{multline}%
}

\usepackage{hyperref}    % Hyperlinks in references
\usepackage[all]{hypcap} % Internal hyperlinks to floats.

\def\lhcb {\mbox{LHCb}\xspace}
\def\ux85 {\mbox{UX85}\xspace}
\def\cern {\mbox{CERN}\xspace}
\def\lhc    {\mbox{LHC}\xspace}

\def\tevatron {Tevatron\xspace}

\ifthenelse{\boolean{uprightparticles}}%
{

 \def\Ppi         {\ensuremath{\uppi}\xspace}

 \def\Pphi        {\ensuremath{\upphi}\xspace}

 \def\PDelta      {\ensuremath{\Delta}\xspace}                 
 \def\PXi      {\ensuremath{\Xi}\xspace}                 
 \def\PLambda      {\ensuremath{\Lambda}\xspace}                 
 \def\PSigma      {\ensuremath{\Sigma}\xspace}                 
 \def\POmega      {\ensuremath{\Omega}\xspace}                 
 \def\PUpsilon      {\ensuremath{\Upsilon}\xspace}                 
 
 \def\PB      {\ensuremath{\mathrm{B}}\xspace}                 
                  
 \def\PD      {\ensuremath{\mathrm{D}}\xspace}

 \def\PH      {\ensuremath{\mathrm{H}}\xspace}

 \def\PK      {\ensuremath{\mathrm{K}}\xspace}

 \def\Pb      {\ensuremath{\mathrm{b}}\xspace}                 
 \def\Pc      {\ensuremath{\mathrm{c}}\xspace}                 
                  
 \def\Pe      {\ensuremath{\mathrm{e}}\xspace}

 \def\Pi      {\ensuremath{\mathrm{i}}\xspace}

 \def\Pp      {\ensuremath{\mathrm{p}}\xspace}

 \def\Ps      {\ensuremath{\mathrm{s}}\xspace}

}
{

 \def\Ppi         {\ensuremath{\pi}\xspace}

 \def\Pphi        {\ensuremath{\phi}\xspace}

 \mathchardef\PDelta="7101
 \mathchardef\PXi="7104
 \mathchardef\PLambda="7103
 \mathchardef\PSigma="7106
 \mathchardef\POmega="710A
 \mathchardef\PUpsilon="7107
                  
 \def\PB      {\ensuremath{B}\xspace}                 
                  
 \def\PD      {\ensuremath{D}\xspace}

 \def\PH      {\ensuremath{H}\xspace}

 \def\PK      {\ensuremath{K}\xspace}

 \def\Pb      {\ensuremath{b}\xspace}                 
 \def\Pc      {\ensuremath{c}\xspace}                 
                  
 \def\Pe      {\ensuremath{e}\xspace}

 \def\Pi      {\ensuremath{i}\xspace}

 \def\Pp      {\ensuremath{p}\xspace}

 \def\Ps      {\ensuremath{s}\xspace}

}

\def\en         {\ensuremath{\Pe^-}\xspace}   % electron negative (\em is taken)
\def\ep         {\ensuremath{\Pe^+}\xspace}
 
\def\epem       {\ensuremath{\Pe^+\Pe^-}\xspace}

\def\squark    {\ensuremath{\Ps}\xspace}

\def\cquark    {\ensuremath{\Pc}\xspace}
\def\cquarkbar {\ensuremath{\overline \cquark}\xspace}
\def\ccbar     {\ensuremath{\cquark\cquarkbar}\xspace}
\def\bquark    {\ensuremath{\Pb}\xspace}

\def\pion  {\ensuremath{\Ppi}\xspace}

\def\pip   {\ensuremath{\pion^+}\xspace}
\def\pim   {\ensuremath{\pion^-}\xspace}

\def\kaon  {\ensuremath{\PK}\xspace}
  \def\Kbar  {\kern 0.2em\overline{\kern -0.2em \PK}{}\xspace}

\def\Kz    {\ensuremath{\kaon^0}\xspace}
\def\Kzb   {\ensuremath{\Kbar^0}\xspace}
\def\KzKzb {\ensuremath{\Kz \kern -0.16em \Kzb}\xspace}
\def\Kp    {\ensuremath{\kaon^+}\xspace}
\def\Km    {\ensuremath{\kaon^-}\xspace}

\def\KpKm  {\ensuremath{\Kp \kern -0.16em \Km}\xspace}
\def\KS    {\ensuremath{\kaon^0_{\rm\scriptscriptstyle S}}\xspace}

  \def\Dbar    {\kern 0.2em\overline{\kern -0.2em \PD}{}\xspace}
\def\D       {\ensuremath{\PD}\xspace}

\def\Dz      {\ensuremath{\D^0}\xspace}
\def\Dzb     {\ensuremath{\Dbar^0}\xspace}
\def\DzDzb   {\ensuremath{\Dz {\kern -0.16em \Dzb}}\xspace}
\def\Dp      {\ensuremath{\D^+}\xspace}
\def\Dm      {\ensuremath{\D^-}\xspace}
\def\Dpm     {\ensuremath{\D^\pm}\xspace}

\def\DpDm    {\ensuremath{\Dp {\kern -0.16em \Dm}}\xspace}

\def\Dstarp  {\ensuremath{\D^{*+}}\xspace}

\def\Dstarpm {\ensuremath{\D^{*\pm}}\xspace}

\def\Ds      {\ensuremath{\D^+_\squark}\xspace}
\def\Dsp     {\ensuremath{\D^+_\squark}\xspace}

\def\Dspm    {\ensuremath{\D^{\pm}_\squark}\xspace}

  \def\Bbar    {\kern 0.18em\overline{\kern -0.18em \PB}{}\xspace}

  \def\Y#1S{\ensuremath{\PUpsilon{(#1S)}}\xspace}% no space before {...}!

\def\proton      {\ensuremath{\Pp}\xspace}
\def\antiproton  {\ensuremath{\overline \proton}\xspace}

\def\Lbar {\ensuremath{\kern 0.1em\overline{\kern -0.1em\PLambda}}\xspace}
\def\Lambdares {\ensuremath{\PLambda}\xspace}

\def\Lc      {\ensuremath{\L^+_\cquark}\xspace}

\def\BF         {{\ensuremath{\cal B}}\xspace}

\def\BR         {\BF}
\newcommand{\decay}[2]{\mbox{\ensuremath{#1\!\to #2}}\xspace}   % {\Pa}{\Pb \Pc}

\def\to                 {\ensuremath{\rightarrow}\xspace}

\newcommand{\Delm}{\mbox{$\Delta m $}\xspace}

\def\AT#1     {\ensuremath{A_{\mathrm{T}}^{#1}}\xspace}           % 2

\def\C#1      {\ensuremath{\mathcal{C}_{#1}}\xspace}                       % 9
\def\Cp#1     {\ensuremath{\mathcal{C}_{#1}^{'}}\xspace}                    % 7
\def\Ceff#1   {\ensuremath{\mathcal{C}_{#1}^{\mathrm{(eff)}}}\xspace}        % 9  
\def\Cpeff#1  {\ensuremath{\mathcal{C}_{#1}^{'\mathrm{(eff)}}}\xspace}       % 7
\def\Ope#1    {\ensuremath{\mathcal{O}_{#1}}\xspace}                       % 2
\def\Opep#1   {\ensuremath{\mathcal{O}_{#1}^{'}}\xspace}                    % 7

\newcommand{\tev}{\ensuremath{\mathrm{\,Te\kern -0.1em V}}\xspace}
\newcommand{\gev}{\ensuremath{\mathrm{\,Ge\kern -0.1em V}}\xspace}
\newcommand{\mev}{\ensuremath{\mathrm{\,Me\kern -0.1em V}}\xspace}
\newcommand{\kev}{\ensuremath{\mathrm{\,ke\kern -0.1em V}}\xspace}
\newcommand{\ev}{\ensuremath{\mathrm{\,e\kern -0.1em V}}\xspace}
\newcommand{\gevc}{\ensuremath{{\mathrm{\,Ge\kern -0.1em V\!/}c}}\xspace}
\newcommand{\mevc}{\ensuremath{{\mathrm{\,Me\kern -0.1em V\!/}c}}\xspace}
\newcommand{\gevcc}{\ensuremath{{\mathrm{\,Ge\kern -0.1em V\!/}c^2}}\xspace}
\newcommand{\gevgevcccc}{\ensuremath{{\mathrm{\,Ge\kern -0.1em V^2\!/}c^4}}\xspace}
\newcommand{\mevcc}{\ensuremath{{\mathrm{\,Me\kern -0.1em V\!/}c^2}}\xspace}

\def\mum  {\ensuremath{\,\upmu\rm m}\xspace}

\def\mub{\ensuremath{\rm \,\upmu b}\xspace}

\def\invnb {\ensuremath{\mbox{\,nb}^{-1}}\xspace}

\newcommand{\stat}{\ensuremath{\mathrm{(stat)}}\xspace}
\newcommand{\syst}{\ensuremath{\mathrm{(syst)}}\xspace}

\newcommand{\chisq}{\ensuremath{\chi^2}\xspace}

\def\gsim{{~\raise.15em\hbox{$>$}\kern-.85em
          \lower.35em\hbox{$\sim$}~}\xspace}
\def\lsim{{~\raise.15em\hbox{$<$}\kern-.85em
          \lower.35em\hbox{$\sim$}~}\xspace}

\def\pt         {\mbox{$p_{\rm T}$}\xspace}

\newcommand{\lum} {\ensuremath{\mathcal{L}}\xspace}
\def\evtgen     {\mbox{\textsc{EvtGen}}\xspace}
\def\pythia     {\mbox{\textsc{Pythia}}\xspace}

\def\geant      {\mbox{\textsc{Geant4}}\xspace}

\def\photos     {\mbox{\textsc{Photos}}\xspace}

\def\tell1  {TELL1\xspace}
\def\ukl1   {UKL1\xspace}

\newcommand{\eg}{\mbox{\itshape e.g.}\xspace}

\newcommand{\etal}{{\slshape et al.\/}\xspace}

\usepackage{cite} % Allows for ranges in citations
\usepackage{mciteplus}

\usepackage{rotating}
\usepackage{subfig}

\def\LHCb {\lhcb}

\def\Lc      {\ensuremath{\Lambdares_\cquark}\xspace}

\def\Lcp     {\ensuremath{\Lambdares_\cquark^+}\xspace}

\def\Lcpm    {\ensuremath{\Lambdares_\cquark^{\pm}}\xspace}

\newcommand{\TeV}{\tev}

\newcommand{\keV}{\keV}

\newcommand{\GeVc}{\gevc}
\newcommand{\MeVc}{\mevc}

\newcommand{\MeVcc}{\mevcc}

\def\mum  {\ensuremath{\,\upmu{\rm m}}\xspace}

\def\mub{\ensuremath{{\rm \,\upmu b}}\xspace}

\def\invnb {\ensuremath{\mbox{\,nb}^{-1}}\xspace}

\def\mubNS       {\ensuremath{{\rm \upmu b}}\xspace}

\newcommand{\dxdy}[2]{\ensuremath{\mathrm{d}\hspace{-0.1em}#1/\mathrm{d}#2}\xspace}
\newcommand{\altdxdy}[2]{\ensuremath{\frac{\mathrm{d}\hspace{-0.1em}#1}{\mathrm{d}\hspace{-0.1em}#2}}\xspace}

\newcommand{\dxdyinline}[2]{\ensuremath{{\mathrm{d}{#1}}/{\mathrm{d}{#2}}}\xspace}

\def\pT         {\pt}

\newcommand{\cquarkto}[1]{\mbox{\ensuremath{\cquark \to #1}}\xspace}

\newcommand{\DzToKmpip}{\decay{\Dz}{\Km \pip}}
\newcommand{\DzToKmpippimpip}{\decay{\Dz}{\Km \pip\pim\pip}}

\newcommand{\DzbToKmpip}{\decay{\Dzb}{\Km \pip}}
\newcommand{\DzDzbToKmpip}{\decay{\ensuremath{(\Dz + \Dzb)}}{\Km \pip}}
\newcommand{\DpToKmpippip}{\decay{\Dp}{\Km \pip\pip}}

\newcommand{\DpTophipip}{\decay{\Dp}{\Pphi(\Km \Kp)\pip}}
\newcommand{\DspTophipip}{\decay{\Dsp}{\Pphi(\Km \Kp) \pip}}

\newcommand{\DspToKmKppip}{\decay{\Dsp}{\Km \Kp \pip}}

\newcommand{\DstarpTopipDz}{\decay{\Dstarp}{\Dz \pip}}
\newcommand{\DstarpTopipDzToKmpip}{\decay{\Dstarp}{\Dz(\Km \pip) \pip}}

\newcommand{\LambdacpTopKmpip}{\decay{\Lcp}{\proton \Km \pip}}

\newcommand{\GeVcNS}{\ensuremath{{\mathrm{Ge\kern -0.1em V\!/}c}}\xspace}

\newcommand{\frag}{\ensuremath{\mathrm{(frag)}}\xspace}

\newcommand{\AGaussExp}{\ensuremath{f_{\mathrm{BG}}}\xspace}

\newcommand{\ptrange}{\mbox{\ensuremath{0 < \pT < 8\GeVc}}\xspace}
\newcommand{\yrange}{\mbox{\ensuremath{2.0 < y < 4.5}}\xspace}

\newcommand{\ip}{\ensuremath{\mathrm{IP}}\xspace}	% Impact parameter
\newcommand{\ipchisq}{\ensuremath{\ip\,\chisq}\xspace}
\newcommand{\logipchisq}{\ensuremath{\log_{10}(\ipchisq)}\xspace}

\newcommand\pythiasix     {\mbox{{\pythia}~6.4}\xspace}

\newcommand{\Tabref}[1]{Table~\ref{#1}}

\newcommand{\figref}[1]{Fig.~\ref{#1}}

\newcommand{\refref}[1]{Ref.~\cite{#1}}

\newcommand{\Tabsref}[2]{\mbox{Tables~\ref{#1}--\ref{#2}}}

\newcommand{\Figsref}[2]{\mbox{Figures~\ref{#1}--\ref{#2}}}
\newcommand{\figsref}[2]{\mbox{Figs.~\ref{#1}--\ref{#2}}}

\newcommand{\xsecstxt}{cross-sections\xspace}

\newcommand{\xsectxtnoun}{cross-section\xspace}
\newcommand{\xsecstxtnoun}{cross-sections\xspace}
\newcommand{\xsectxtadj}{cross-section\xspace}
\newcommand{\Xsectxtnoun}{Cross-section\xspace}
\newcommand{\Xsecstxtnoun}{Cross-sections\xspace}
\newcommand{\Xsectxtadj}{Cross-section\xspace}

\newcommand{\FONLLtext}{Fixed Order Next to Leading Logarithm\xspace}
\newcommand{\GMVFNStext}{Generalized Mass Variable Flavour Number Scheme\xspace}

\newcommand{\lumifivepmerr}{\ensuremath{1.9 \pm 0.1\invnb}\xspace}

\newcommand{\lumieightpmerr}{\ensuremath{13.1 \pm 0.5\invnb}\xspace}

\newcommand{\lumipmerr}{\ensuremath{15.0 \pm 0.5\invnb}\xspace}

\newcommand{\lumitwosigfig}{\ensuremath{15\invnb}\xspace}

\newcommand{\totccbarxsec}{\ensuremath{1419 \pm 12\,\stat \pm 116\,\syst \pm 65\,\frag \mub}\xspace}

\usepackage[defaultlines=3,all]{nowidow}

\usepackage{afterpage}

\begin{document}

\renewcommand{\thefootnote}{\fnsymbol{footnote}}
\setcounter{footnote}{1}

%% Title
%% ========================================================================== %%
% $Id: title-LHCb-PAPER.tex 31049 2013-02-06 16:07:01Z spradlin $
% ===============================================================================
% Purpose: LHCb-PAPER journal paper title page template
% Author: 
% Created on: 2010-09-25
% ===============================================================================

%%%%%%%%%%%%%%%%%%%%%%%%%
%%%%%  TITLE PAGE  %%%%%%
%%%%%%%%%%%%%%%%%%%%%%%%%
\begin{titlepage}
\pagenumbering{roman}

% Header ---------------------------------------------------
\vspace*{-1.5cm}
\centerline{\large EUROPEAN ORGANIZATION FOR NUCLEAR RESEARCH (CERN)}
\vspace*{1.5cm}
\hspace*{-0.5cm}
\begin{tabular*}{\linewidth}{lc@{\extracolsep{\fill}}r}
\ifthenelse{\boolean{pdflatex}}% Logo format choice
{\vspace*{-2.7cm}\mbox{\!\!\!\includegraphics[width=.14\textwidth]{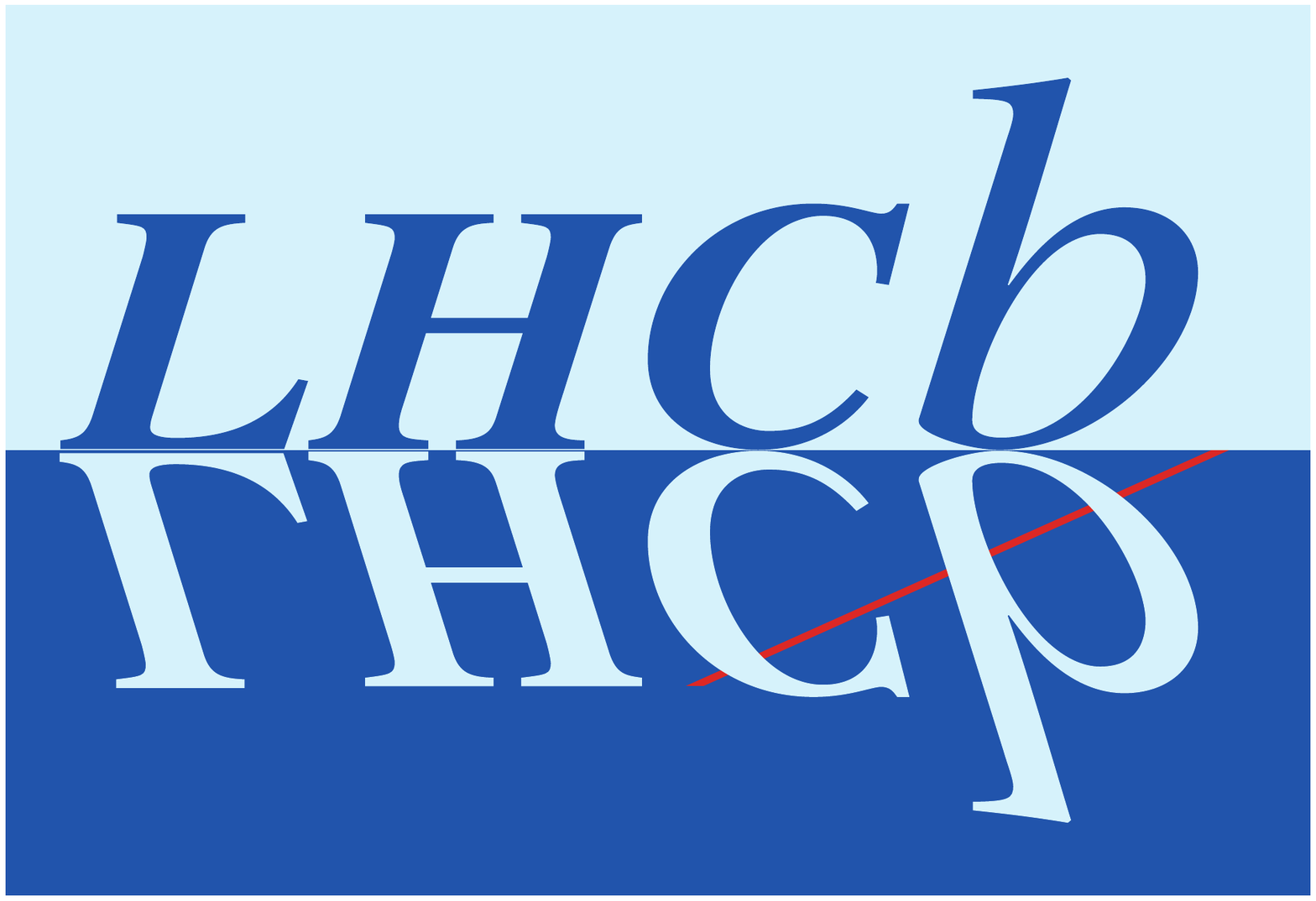}} & &}%
{\vspace*{-1.2cm}\mbox{\!\!\!\includegraphics[width=.12\textwidth]{figs/lhcb-logo.eps}} & &}%
\\
 & & CERN-PH-EP-2013-009 \\  % ID 
 & & LHCb-PAPER-2012-041 \\  % ID 
 & & 12 February 2013 \\ % Date of collaboration approval to submit.
 & & \\
% not in paper \hline
\end{tabular*}

%\vspace*{4.0cm}
\vspace*{3.0cm}

% Title --------------------------------------------------
{\bf\boldmath\huge
\begin{center}
  Prompt charm production in $\proton\proton$~collisions at $\sqrt{s}= 7\TeV$
\end{center}
}

%\vspace*{2.0cm}
\vspace*{1.0cm}

% Authors -------------------------------------------------
\begin{center}
The LHCb collaboration\footnote{Authors are listed on the following pages.}
\end{center}

\vspace{\fill}

% Abstract -----------------------------------------------
\begin{abstract}
  \noindent
    Charm production at the \lhc in $\proton\proton$
    collisions at \mbox{$\sqrt{s}= 7\TeV$} is studied with the \LHCb detector.
    The decays \DzToKmpip, \DpToKmpippip, \DstarpTopipDzToKmpip, \DspTophipip,
    \LambdacpTopKmpip, and their charge conjugates are analysed in a data set
    corresponding to an integrated luminosity of \lumitwosigfig.
    Differential \xsecstxtnoun \dxdyinline{\ensuremath{\sigma}}{\pT} are
    measured for prompt production of the five charmed hadron species
    in bins of transverse momentum and rapidity in the region
    \ptrange and \yrange.
    Theoretical predictions are compared to the measured differential
    \xsecstxtnoun.
    The integrated \xsecstxtnoun of the charm hadrons are computed in the
    above \pT-$y$ range, and their ratios are reported.
    A combination of the five integrated \xsectxtadj measurements gives
    \begin{equation*}
      \sigma(\ccbar)_{p_{\mathrm{T}} < 8\mathrm{\,Ge\kern -0.1em V\!/}c, \,2.0 < y < 4.5}
        =  \totccbarxsec,
    \end{equation*}
    where the uncertainties are statistical, systematic, and due to the
    fragmentation functions.
\end{abstract}

%\vspace*{2.0cm}
\vspace*{1.0cm}

\begin{center}
  Submitted to Nuclear Physics B
\end{center}

\vspace{\fill}

{\footnotesize
\centerline{\copyright~CERN on behalf of the \lhcb collaboration, license \href{http://creativecommons.org/licenses/by/3.0/}{CC-BY-3.0}.}}
\vspace*{2mm}

\end{titlepage}

%%%%%%%%%%%%%%%%%%%%%%%%%%%%%%%%
%%%%%  EOD OF TITLE PAGE  %%%%%%
%%%%%%%%%%%%%%%%%%%%%%%%%%%%%%%%

%  empty page follows the title page ----
\newpage
\setcounter{page}{2}
\mbox{~}
\newpage

% Author List ----------------------------
%  You need to get a new author list!
%%%%%%%%%%%%%%%%%%%%%%%%%%%%%%%%%%%%%%%%%%
\centerline{\large\bf LHCb collaboration}
\begin{flushleft}
\small
R.~Aaij$^{38}$, 
C.~Abellan~Beteta$^{33,n}$, 
A.~Adametz$^{11}$, 
B.~Adeva$^{34}$, 
M.~Adinolfi$^{43}$, 
C.~Adrover$^{6}$, 
A.~Affolder$^{49}$, 
Z.~Ajaltouni$^{5}$, 
J.~Albrecht$^{9}$, 
F.~Alessio$^{35}$
% ...

{\footnotesize \it
$ ^{1}$Centro Brasileiro de Pesquisas F\'{i}sicas (CBPF), Rio de Janeiro, Brazil\\
% ...
$ ^{5}$Clermont Universit\'{e}, Universit\'{e} Blaise Pascal, CNRS/IN2P3, LPC, Clermont-Ferrand, France\\
$ ^{6}$CPPM, Aix-Marseille Universit\'{e}, CNRS/IN2P3, Marseille, France\\
% ...
$ ^{9}$Fakult\"{a}t Physik, Technische Universit\"{a}t Dortmund, Dortmund, Germany\\
% ...
$ ^{11}$Physikalisches Institut, Ruprecht-Karls-Universit\"{a}t Heidelberg, Heidelberg, Germany\\
% ...
$ ^{33}$Universitat de Barcelona, Barcelona, Spain\\
$ ^{34}$Universidad de Santiago de Compostela, Santiago de Compostela, Spain\\
$ ^{35}$European Organization for Nuclear Research (CERN), Geneva, Switzerland\\
% ...
$ ^{38}$Nikhef National Institute for Subatomic Physics, Amsterdam, The Netherlands\\
% ...
$ ^{43}$H.H. Wills Physics Laboratory, University of Bristol, Bristol, United Kingdom\\
% ...
$ ^{49}$Oliver Lodge Laboratory, University of Liverpool, Liverpool, United Kingdom\\
% ...
$ ^{n}$LIFAELS, La Salle, Universitat Ramon Llull, Barcelona, Spain\\
% ...
}
\end{flushleft}
%%%%%%%%%%%%%%%%%%%%%%%%%%%%%%%%%%%%%%%%%%

\cleardoublepage

%% Reset foonote counter.
%% -------------------------------------------------------------------------- %%
\renewcommand{\thefootnote}{\arabic{footnote}}
\setcounter{footnote}{0}

%% Table of Content
%% ========================================================================== %%
%\tableofcontents
%\clearpage
%\listoffigures
%\clearpage
%\listoftables
%\cleardoublepage

%% Main text %%%%%%%%%
%% ========================================================================== %%
\pagestyle{plain}       % restore page numbers for the main text
\setcounter{page}{1}
\pagenumbering{arabic}

%% Line numbering.
%% -------------------------------------------------------------------------- %%
%\linenumbers

%% $Id: introduction.tex 31037 2013-02-06 13:38:55Z spradlin $
%% ====================================================================== %%
\section{Introduction}
\label{sec:intro}

Measurements of the production \xsecstxtnoun of charmed hadrons
test the predictions of quantum
chromodynamic (QCD) fragmentation and hadronisation
models.
Perturbative calculations of charmed hadron production \xsecstxtnoun at
next-to-leading order using the \GMVFNStext
(GMVFNS)~\cite{Kniehl:2004fy,Kniehl:2005gz,Kniehl:2005ej,Kneesch:2007ey,Kniehl:2009ar,Kniehl:2012ti}
and at fixed order with next-to-leading-log resummation
(FONLL)~\cite{Cacciari:1998it,Cacciari:2003zu,Cacciari:2005uk,Cacciari:2012ny}
reproduce the \xsecstxtnoun measured in the central rapidity region
($|y| \le 1$) in $\proton\antiproton$ collisions at $\sqrt{s} = 1.97\TeV$ at
the Fermilab \tevatron collider~\cite{Acosta:2003ax} and the
\xsecstxtnoun measured in the central rapidity region ($|y| < 0.5$) in
$\proton\proton$ collisions at $\sqrt{s} = 2.96\TeV$~\cite{Abelev:2012sx}
and at $\sqrt{s} = 7\TeV$~\cite{ALICE:2012ik,ALICE:2011aa} at the \cern
Large Hadron Collider (\lhc).
The \LHCb detector at the \lhc provides unique access to the forward rapidity
region at these energies with a detector that is tailored for flavour physics. 
This paper presents measurements with the \LHCb detector of
\Dz, \Dp, \Dsp, \Dstarp, and \Lcp
production in the forward rapidity region \yrange
in $\proton\proton$ collisions at a centre-of-mass energy of $7\TeV$.
Throughout this article, references to specific decay modes or
specific charmed hadrons also imply the charge conjugate mode.
The measurements are based on \lumitwosigfig of $\proton\proton$
collisions recorded with the \LHCb detector in 2010 with approximately
$1.1$ visible interactions per triggered bunch crossing.

Charmed hadrons may be produced at the $\proton\proton$ collision point
either directly or as feed-down from the instantaneous decays of excited charm
resonances.
They may also be produced in decays of \mbox{\bquark-hadrons}.
In this paper, the first two sources (direct production and feed-down) are
referred to as \textit{prompt}.
Charmed particles from \mbox{\bquark-hadron} decays are called
\textit{secondary} charmed hadrons.
The measurements described here are the production \xsecstxtnoun of prompt
charmed hadrons.
Secondary charmed hadrons are treated as backgrounds.
No attempt is made to distinguish between the two sources of prompt charmed
hadrons.

   % \label{sec:intro}

%% $Id: $
%% ====================================================================== %%
\section{Experimental conditions}
\label{sec:detector}

The \lhcb detector~\cite{Alves:2008zz} is a single-arm forward
spectrometer covering the \mbox{pseudorapidity} range \mbox{$2<\eta <5$},
designed for the study of particles containing \bquark or \cquark quarks.
The detector includes a high precision tracking system consisting of a
silicon-strip vertex detector surrounding the $\proton\proton$ interaction
region, a large-area silicon-strip detector located upstream of a dipole
magnet with a bending power of about $4{\rm\,Tm}$, and three stations
of silicon-strip detectors and straw drift-tubes placed
downstream.
The combined tracking system has a momentum resolution
($\Delta p/p$) that varies from 0.4\% at 5\gevc to 0.6\% at 100\gevc
and an impact parameter (\ip) resolution of 20\mum for tracks with high
transverse momentum.
Charged hadrons are identified using two
ring-imaging Cherenkov detectors.
Photon, electron, and hadron candidates are identified by a calorimeter system
consisting of scintillating-pad and pre-shower detectors, an electromagnetic
calorimeter, and a hadronic calorimeter.
Muons are identified by a system composed of alternating layers of iron
and multiwire proportional chambers.
The trigger consists of a hardware stage, based on information from the
calorimeter and muon systems, followed by a software stage that applies a
full event reconstruction.

During the considered data taking period, the rate of bunch crossings at
the \LHCb interaction point was sufficiently small that the software stage
of the trigger could process all bunch crossings.
Candidate events passed through the hardware stage of the trigger without
filtering.
The software stage of the trigger accepted bunch crossings for which at least
one track was reconstructed in either the silicon-strip vertex detector or the
downstream tracking stations.
The sample is divided into two periods of data collection.
In the first \lumifivepmerr all bunch crossings satisfying these
criteria were retained.
In the subsequent \lumieightpmerr the trigger retention rate was limited to
a randomly selected $(24.0 \pm 0.2)\%$ of all bunch crossings.

For simulated events, $\proton\proton$ collisions are generated using
\pythia~6.4~\cite{Sjostrand:2006za} with a specific \lhcb
configuration~\cite{LHCb-PROC-2010-056} that employs the CTEQ6L1 parton
densities~\cite{Pumplin:2002vw}.
Decays of hadronic particles are described by \evtgen~\cite{Lange:2001uf} in
which final state radiation is generated using \photos~\cite{Golonka:2005pn}.
The interaction of the generated particles with the detector and its
response are implemented using the \geant
toolkit~\cite{Allison:2006ve, *Agostinelli:2002hh} as described in
\refref{LHCb-PROC-2011-006}.

       % \label{sec:detector}

\section{Analysis strategy}
\label{sec:analysis}

The analysis is based on fully reconstructed decays of charmed hadrons
in the following decay modes:
\DzToKmpip, \DpToKmpippip, \DstarpTopipDzToKmpip, \DspTophipip, and
\LambdacpTopKmpip.
Formally, the \DzToKmpip sample contains the sum of the Cabibbo-favoured decays
\DzToKmpip and the doubly Cabibbo-suppressed decays \DzbToKmpip.
For simplicity, we will refer to the combined sample by its dominant component.

The measurements are performed in two-dimensional bins of the transverse
momentum (\pT) and rapidity ($y$) of the reconstructed hadrons, measured with
respect to the beam axis in the $\proton\proton$ centre-of-mass (CM) frame.
For the \Dz, \Dp, \Dstarp, and \Dsp measurements, we use eight bins of uniform
width in the range \ptrange and five bins of uniform width in the range
\yrange.
For the \Lcp measurement, we partition the data in two ways: six uniform \pT
bins in \mbox{$2 < \pT < 8\GeVc$} with a single \yrange bin
and a single \mbox{$2 < \pT < 8\GeVc$} bin with
five uniform $y$ bins in \yrange.

%% Description of selection
%% ------------------------------------------------------------ %%
\subsection{Selection criteria}
\label{sec:analysis:sel}

The selection criteria were tuned independently for each decay.
The same selection criteria are used for \DzToKmpip candidates in the \Dz
and \Dstarp \xsectxtadj measurements.
We use only events that have at least one reconstructed primary interaction
vertex (PV).
Each final state kaon, pion, or proton candidate used in the reconstruction
of a \Dz, \Dp, \Dsp, or \Lcp candidate
must be positively identified.
Because of the relatively long lifetimes of the \Dz, \Dp, \Dsp, and \Lcp
hadrons, the
trajectories of their decay products will not, in general, point directly back
to the PV at which the charmed hadron was produced.
To exploit this feature, the selections for these decays require that
each final state candidate has a minimum impact
parameter \chisq (\ipchisq) with respect to the PV\@.
The \ipchisq is defined as the difference between the \chisq of the PV
reconstructed with and without the considered particle.
For the \Dz and \Lcp reconstruction, a common \ipchisq requirement is imposed
on all final state particles.
For the \Dp and \Dsp candidates, progressively stricter limits are used for the
three daughters.
Final-state decay products of
charmed hadrons have transverse momenta that are generally larger than those
of stable charged particles produced at the PV\@.
Applying lower limits on the \pT of the final state tracks suppresses
combinatorial backgrounds in the selections of
\Dz, \Dp, and \Lcp samples.

\begin{figure}[p]
      \centering
      \subfloat{\label{fig:Dz:m:m}%
        \includegraphics[width=0.495\textwidth]{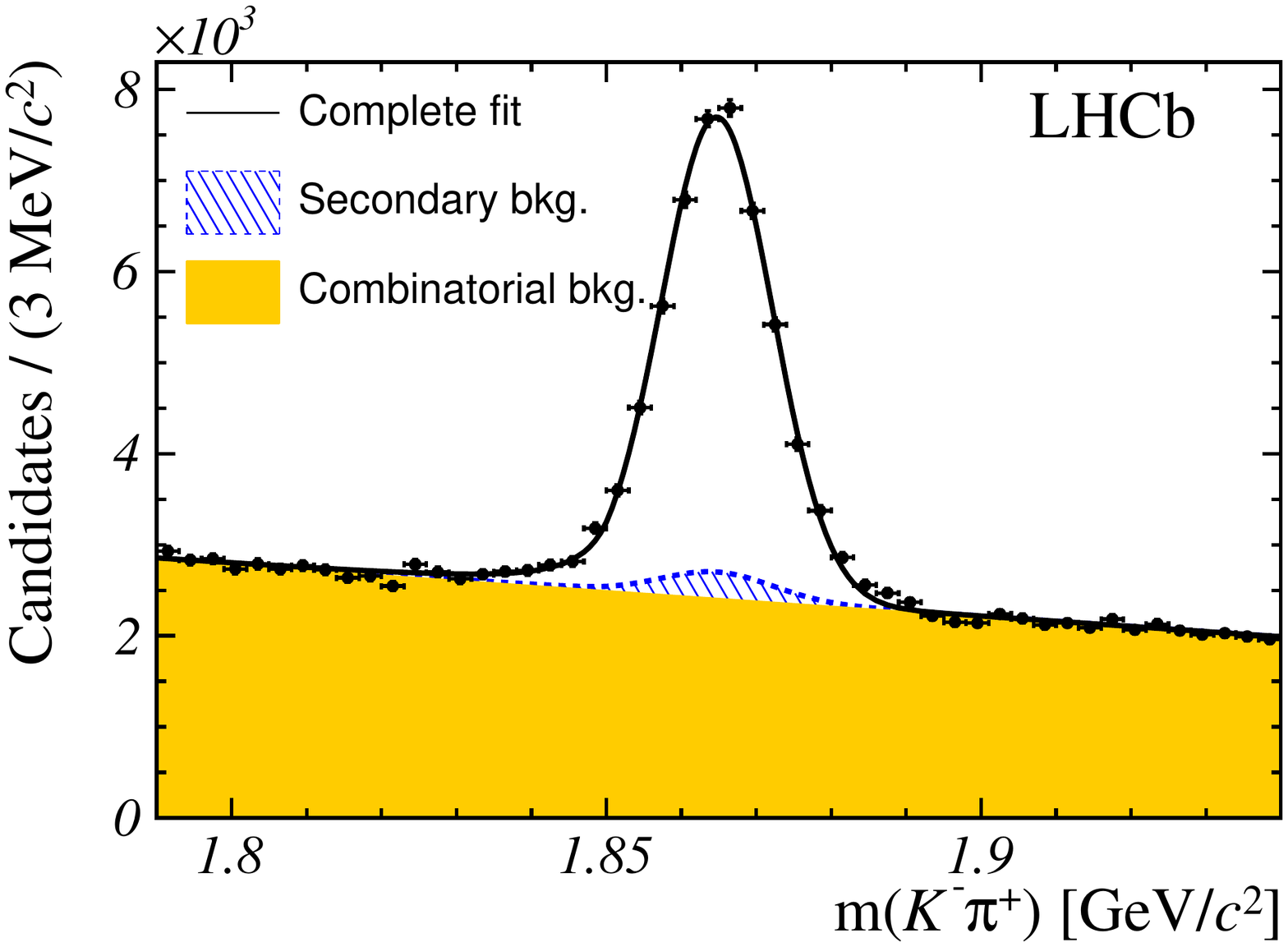}%
        \makebox[0cm][r]{\raisebox{0.19\textheight}[0cm]{\protect\subref{fig:Dz:m:m}}\hspace{0.04\textwidth}}
      }%
      \subfloat{\label{fig:Dz:m:ip}%
        \includegraphics[width=0.495\textwidth]{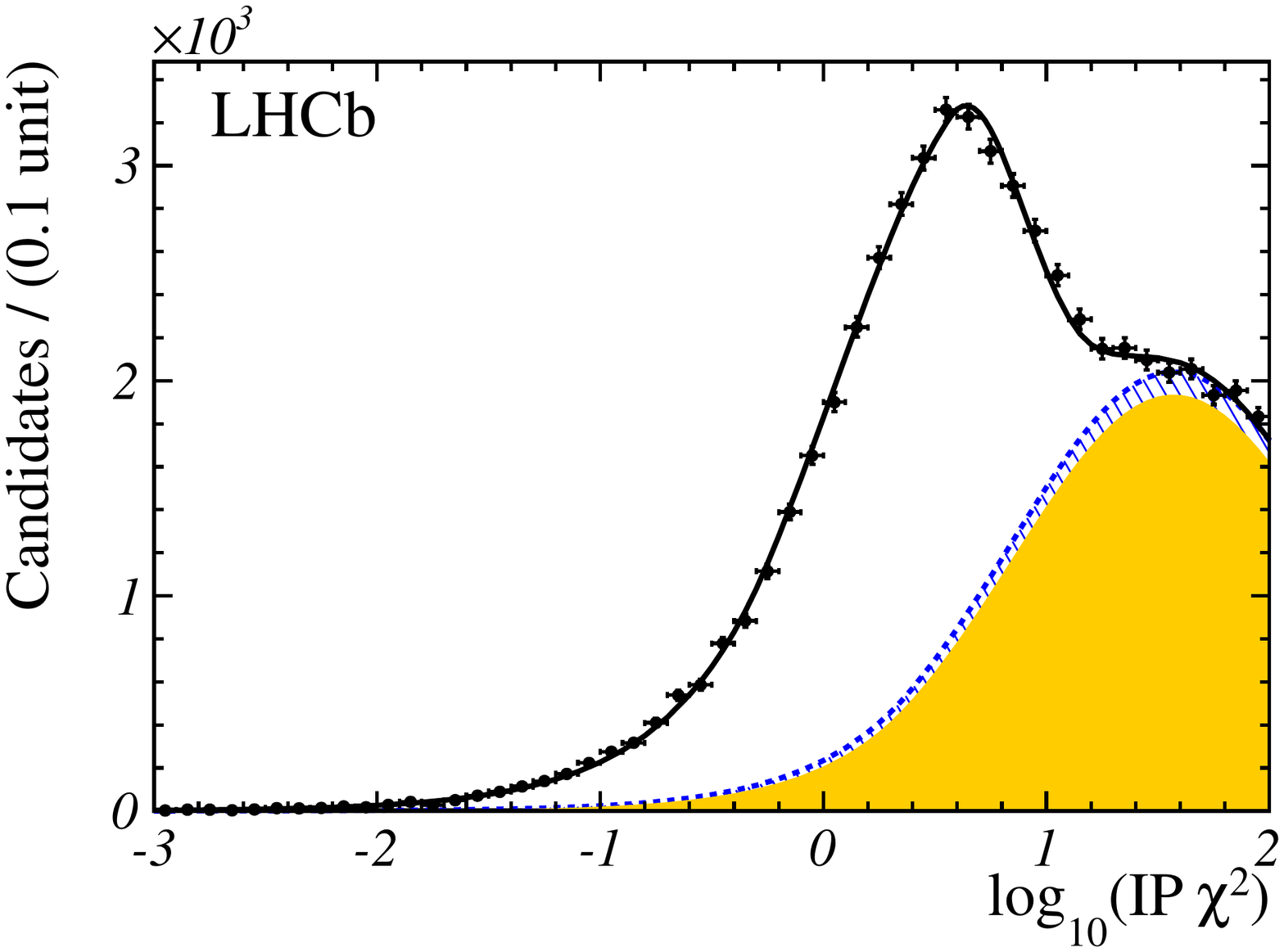}%
        \makebox[0cm][r]{\raisebox{0.19\textheight}[0cm]{\protect\subref{fig:Dz:m:ip}}\hspace{0.37\textwidth}}
      }

      \subfloat{\label{fig:Dp:m:m}%
        \includegraphics*[width=0.495\textwidth]{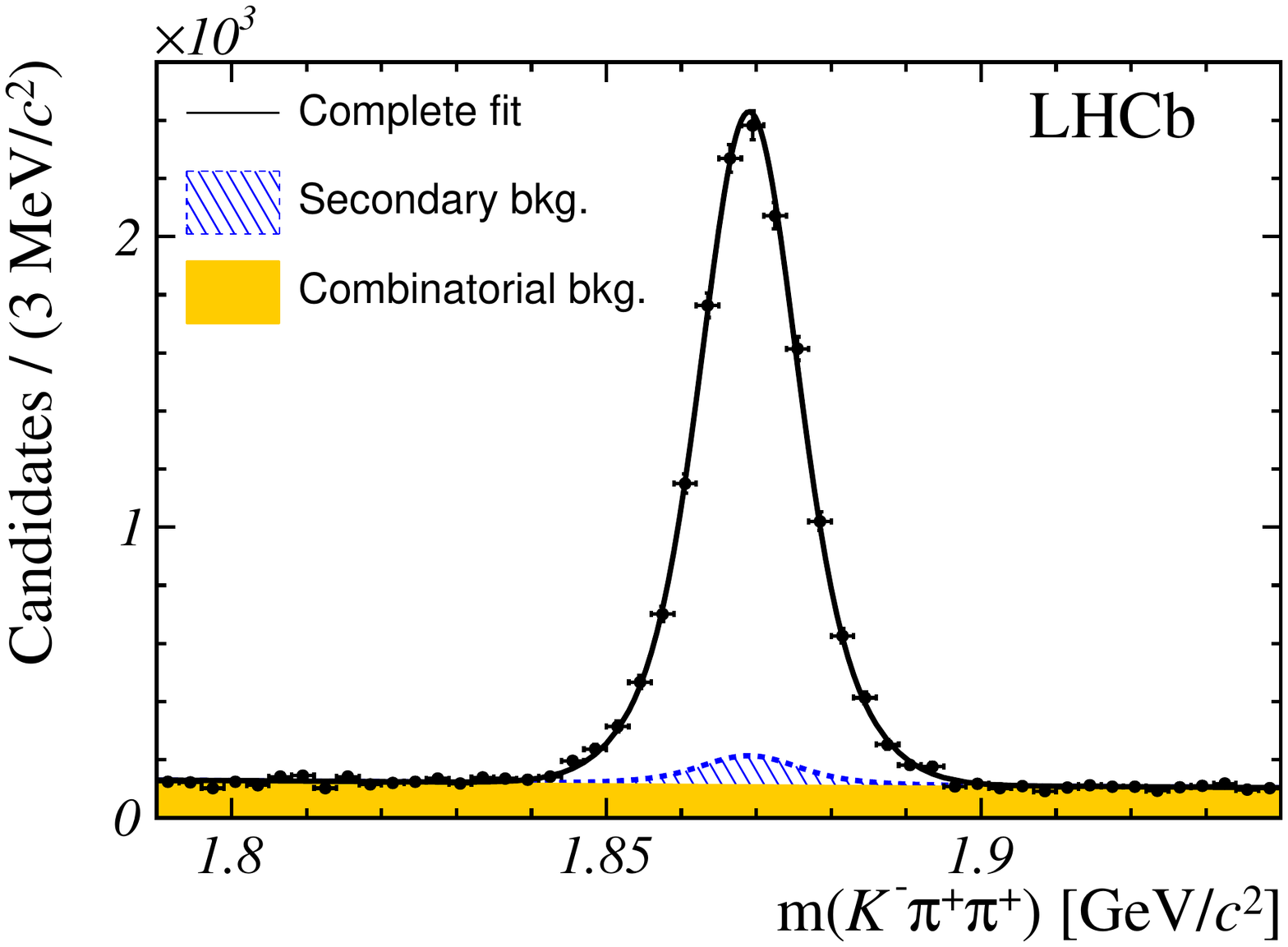}%
        \makebox[0cm][r]{\raisebox{0.19\textheight}[0cm]{\protect\subref{fig:Dp:m:m}}\hspace{0.04\textwidth}}
      }%
      \subfloat{\label{fig:Dp:m:ip}%
        \includegraphics*[width=0.495\textwidth]{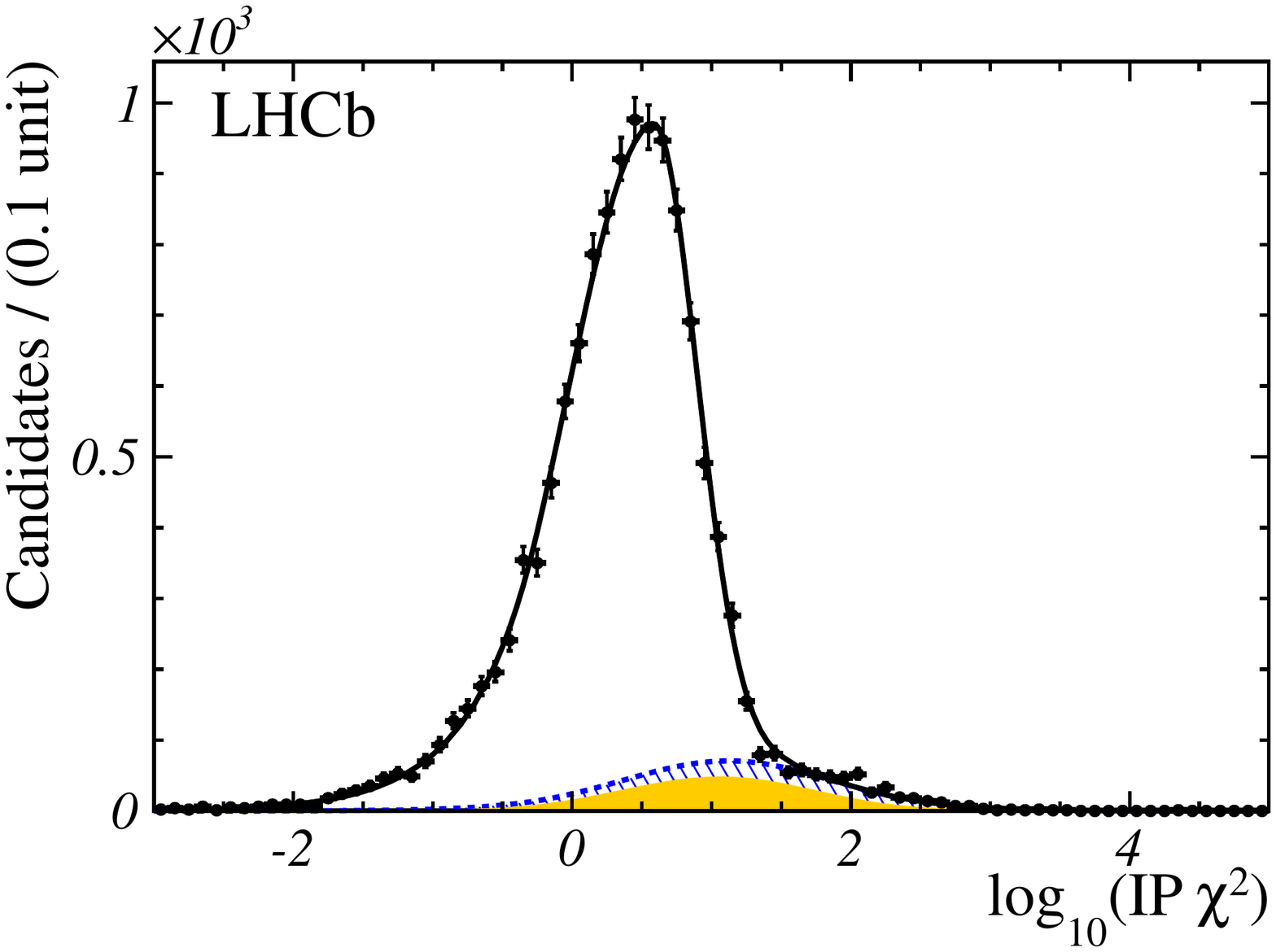}%
        \makebox[0cm][r]{\raisebox{0.19\textheight}[0cm]{\protect\subref{fig:Dp:m:ip}}\hspace{0.37\textwidth}}
      }

      \caption[\DzToKmpip and \DpToKmpippip mass and \logipchisq distributions]{
        \small Mass and \logipchisq distributions for selected \DzToKmpip
        and \DpToKmpippip candidates showing
        \protect\subref{fig:Dz:m:m} the masses of the \Dz candidates,
        \protect\subref{fig:Dz:m:ip} the \logipchisq distribution of \Dz
        candidates for a mass window of \mbox{$\pm 16\MeVcc$} (approximately
        \mbox{$\pm 2\sigma$}) around the fitted $m(\Km\pip)$ peak,
        \protect\subref{fig:Dp:m:m} the masses of the \Dp candidates,
        and \protect\subref{fig:Dp:m:ip} the \logipchisq distribution of
        \Dp candidates for a mass window of \mbox{$\pm 11\MeVcc$}
        (approximately \mbox{$\pm 2\sigma$}) around the fitted
        $m(\Km\pip\pip)$ peak.
        Projections of likelihood fits to the full data samples are shown
        with components as indicated in the legends.
      \label{fig:Dz:m}
      \label{fig:Dp:m}}
    \end{figure}

The selections of candidate charmed hadron decays are further refined by
studying properties of the combinations of the selected final state
particles.
Candidate \DspTophipip decays are required to have a $\Km\Kp$ invariant mass
within \mbox{$\pm 20\MeVcc$} of the $\Pphi(1020)$ mass~\cite{PDG2012}.
The decay products for each candidate charmed hadron must be
consistent with originating from a common vertex with a good quality fit.
The significant lifetimes of \Dz, \Dp, \Dsp, and \Lcp hadrons are
exploited by requiring that the fitted decay vertexes are significantly
displaced from the PV\@.
The trajectory of a prompt charmed hadron should point back to the PV in
which it was produced.
For \Dz candidates this is exploited as a requirement that
\mbox{$\ipchisq < 100$}.
For \Dz decays, we use one additional discriminating variable:  the angle
between the momentum of the \Dz candidate in the laboratory frame and
the momentum of the pion candidate from its decay evaluated in the \Dz rest
frame.
The cosine of this angle has a flat distribution for \Dz decays but
peaks strongly in the forward direction for combinatorial backgrounds.
Candidate \Dstarp decays are reconstructed from \Dz and
\textit{slow} pion candidates.
\Figsref{fig:Dz:m}{fig:Lcp:m} show the invariant mass distributions
and the \logipchisq distributions of the selected charmed hadron candidates.

    \begin{figure}[p]
      \centering
      \subfloat{\label{fig:Dstarp:m:mkpi}%
        \includegraphics[width=0.495\textwidth]{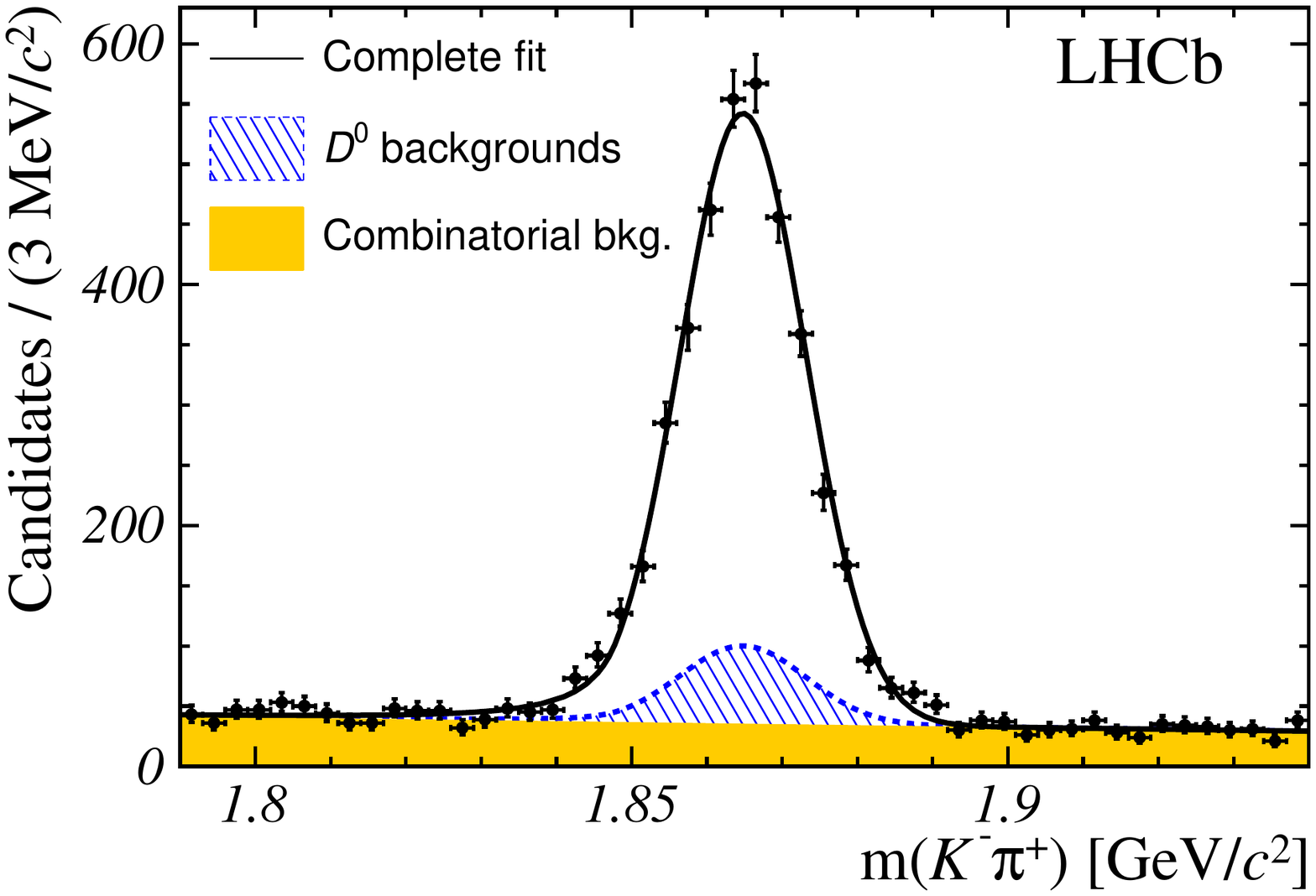}%
        \makebox[0cm][r]{\raisebox{0.19\textheight}[0cm]{\protect\subref{fig:Dstarp:m:mkpi}}\hspace{0.04\textwidth}}
      }%
      \subfloat{\label{fig:Dstarp:m:delm}%
        \includegraphics[width=0.495\textwidth]{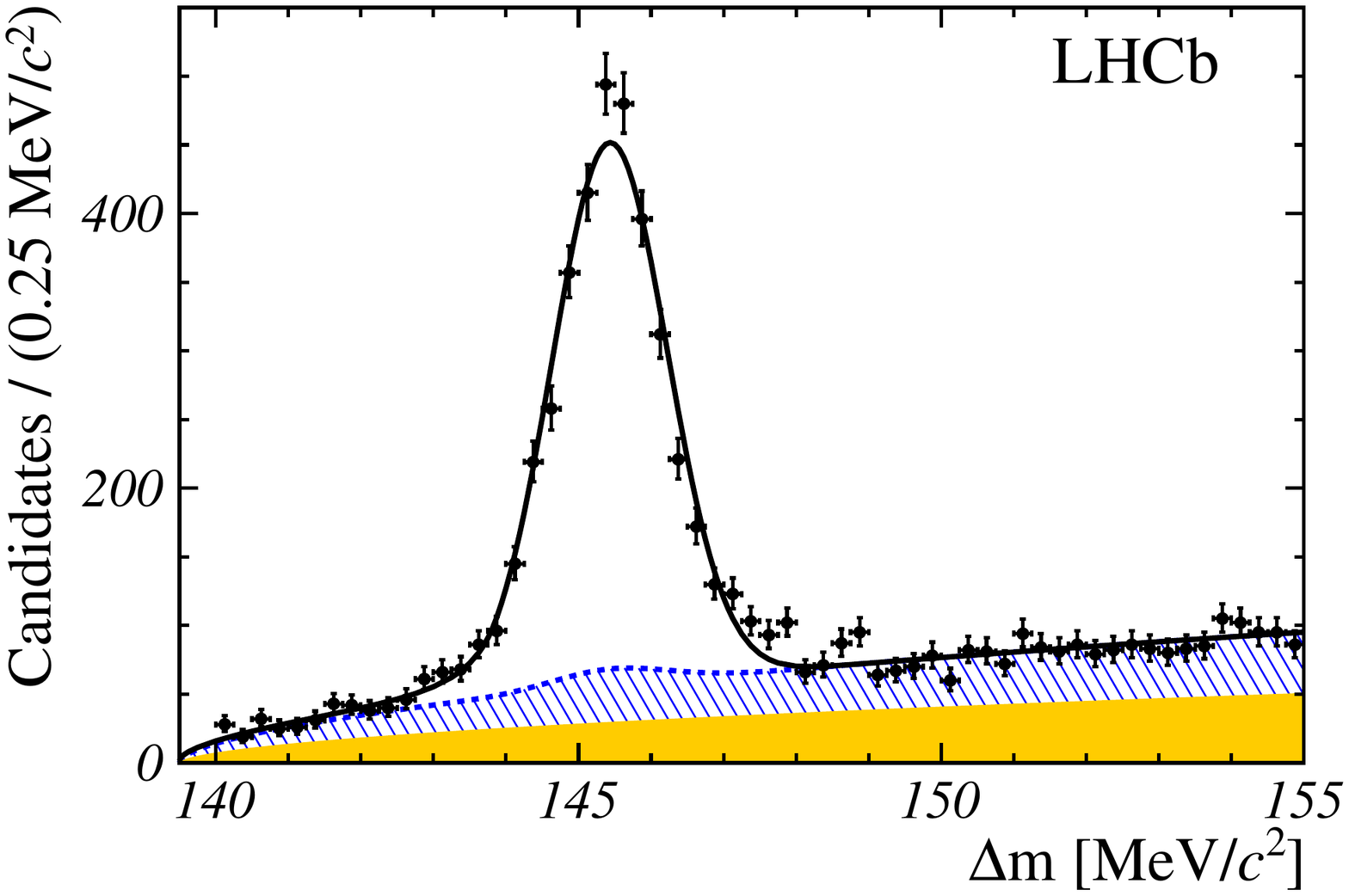}%
        \makebox[0cm][r]{\raisebox{0.19\textheight}[0cm]{\protect\subref{fig:Dstarp:m:delm}}\hspace{0.04\textwidth}}
      }\\

      \subfloat{\label{fig:Dstarp:m:ip}%
        \includegraphics[width=0.495\textwidth]{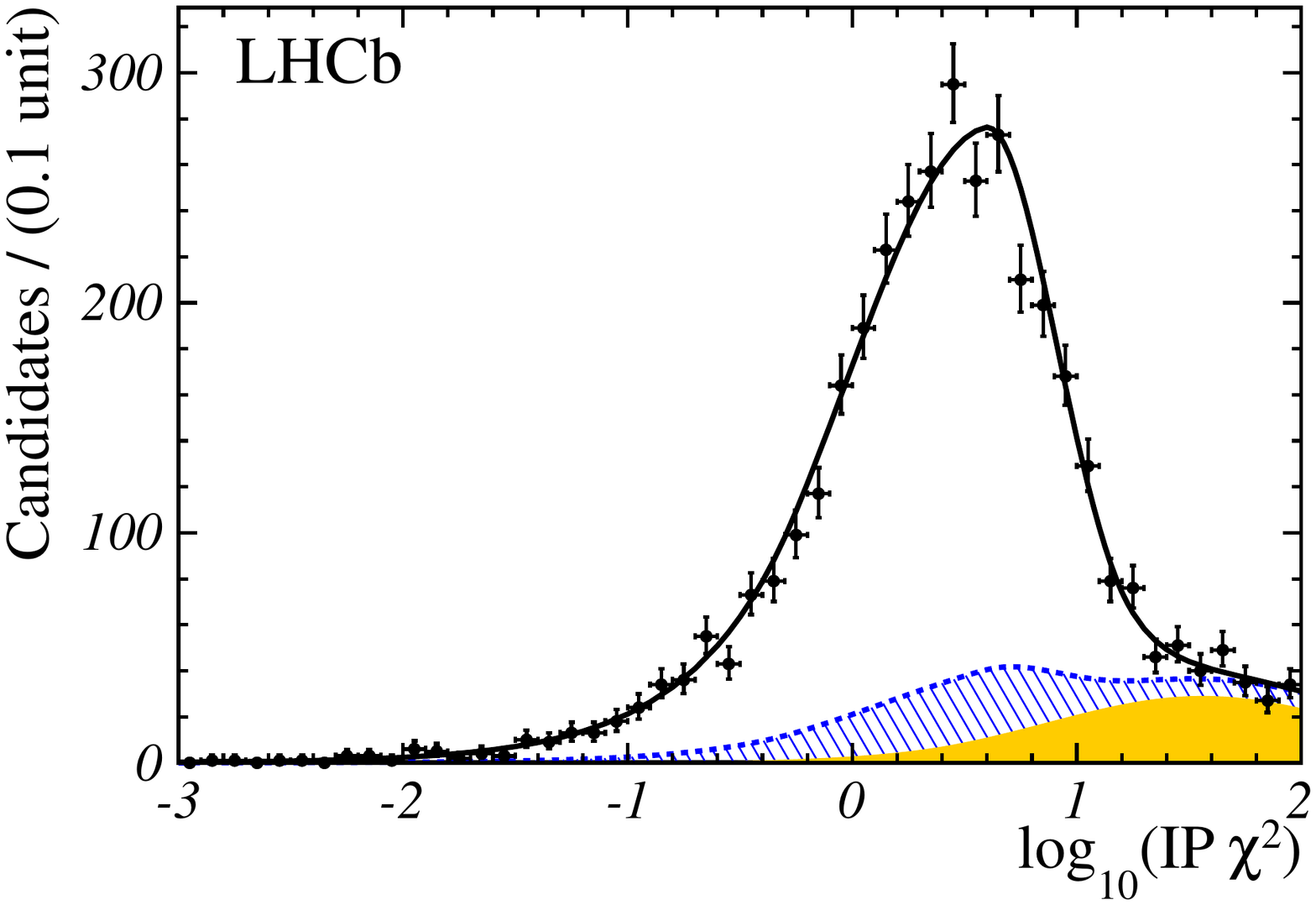}%
        \makebox[0cm][r]{\raisebox{0.19\textheight}[0cm]{\protect\subref{fig:Dstarp:m:ip}}\hspace{0.37\textwidth}}
      }

      \caption[\DstarpTopipDz, \DzToKmpip mass and \logipchisq distributions]{
        \small Mass and \logipchisq distributions
        for selected \DstarpTopipDzToKmpip candidates showing
        \protect\subref{fig:Dstarp:m:mkpi} the masses of the \Dz candidates for
        a window of \mbox{$\pm 1.6\MeVcc$} (approximately $\pm 2\sigma$) around
        the fitted \Delm peak,
        \protect\subref{fig:Dstarp:m:delm} the differences between the \Dstarp
        and \Dz candidate masses for a mass window of \mbox{$\pm 16\MeVcc$}
        (approximately $\pm 2\sigma$) around the fitted $m(\Km\pip)$ peak,
        and \protect\subref{fig:Dstarp:m:ip} the \logipchisq distribution of
        the \Dz candidate for a mass signal box of $\pm 16\MeVcc$ around the
        fitted $m(\Km\pip)$ peak and $\pm 1.6\MeVcc$ around the fitted
        \Delm peak.
        Projections of a likelihood fit to the full data sample are shown
        with components as indicated in the legend.
        The `\Dz backgrounds' component is the sum of the secondary,
        prompt random slow pion, and secondary random slow pion backgrounds.
      \label{fig:Dstarp:m}}
    \end{figure}

    \begin{figure}[p]
      \centering
      \subfloat{\label{fig:Ds:m:m}%
        \includegraphics[width=0.495\textwidth]{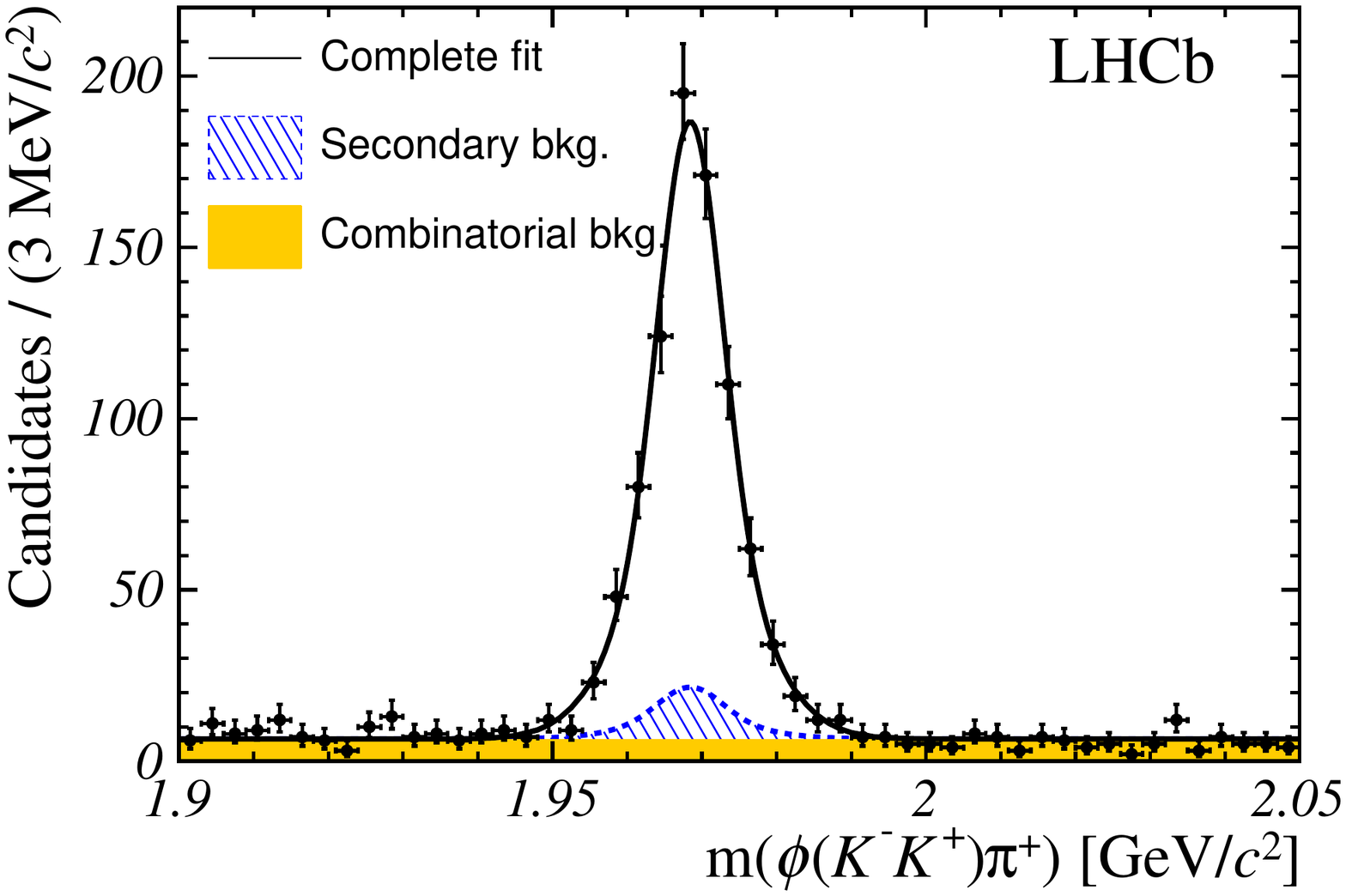}%
        \makebox[0cm][r]{\raisebox{0.19\textheight}[0cm]{\protect\subref{fig:Ds:m:m}}\hspace{0.04\textwidth}}
      }%
      \subfloat{\label{fig:Ds:m:ip}%
        \includegraphics[width=0.495\textwidth]{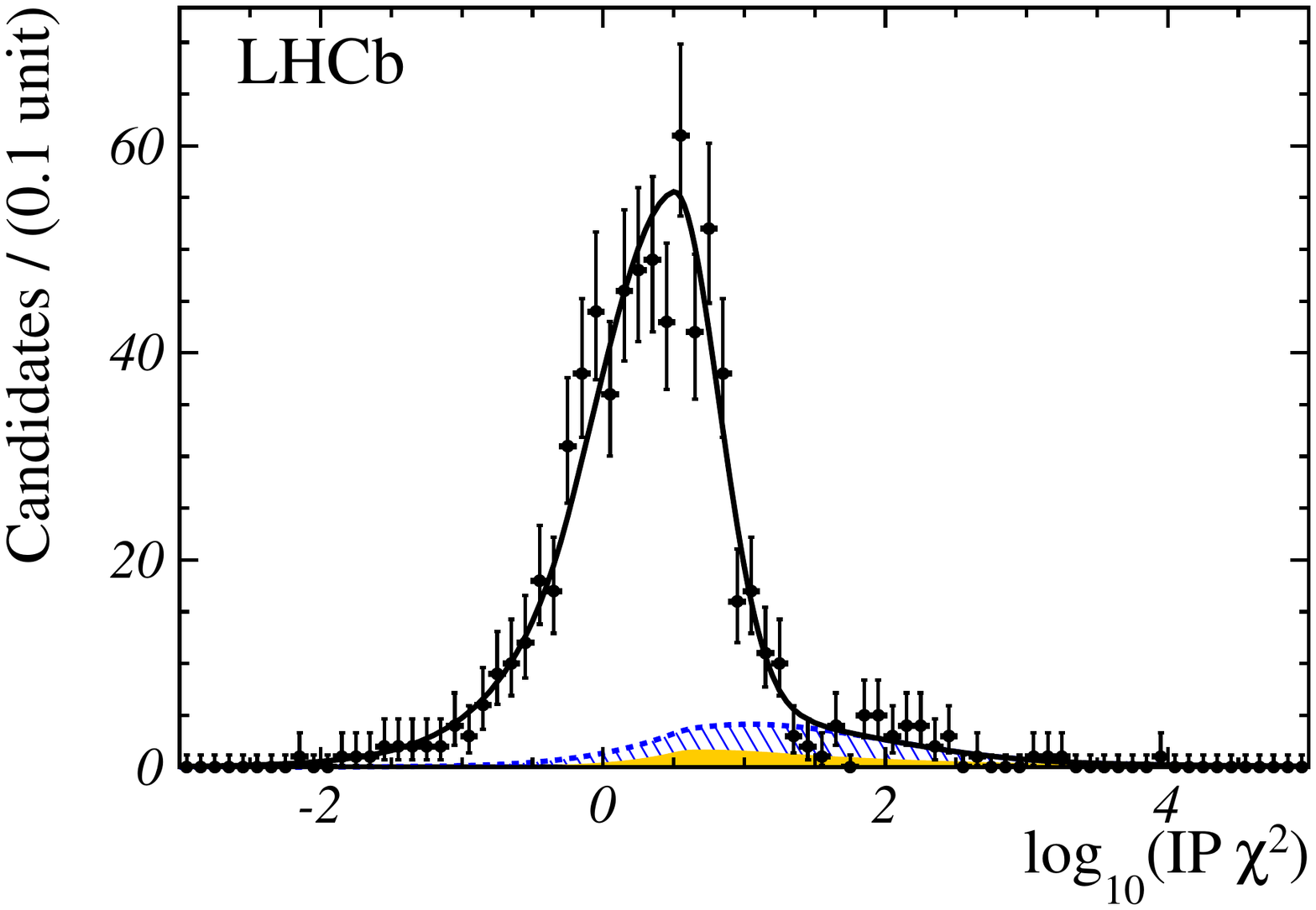}%
        \makebox[0cm][r]{\raisebox{0.19\textheight}[0cm]{\protect\subref{fig:Ds:m:ip}}\hspace{0.37\textwidth}}
      }

      \subfloat{\label{fig:Lcp:m:m}%
        \includegraphics[width=0.495\textwidth]{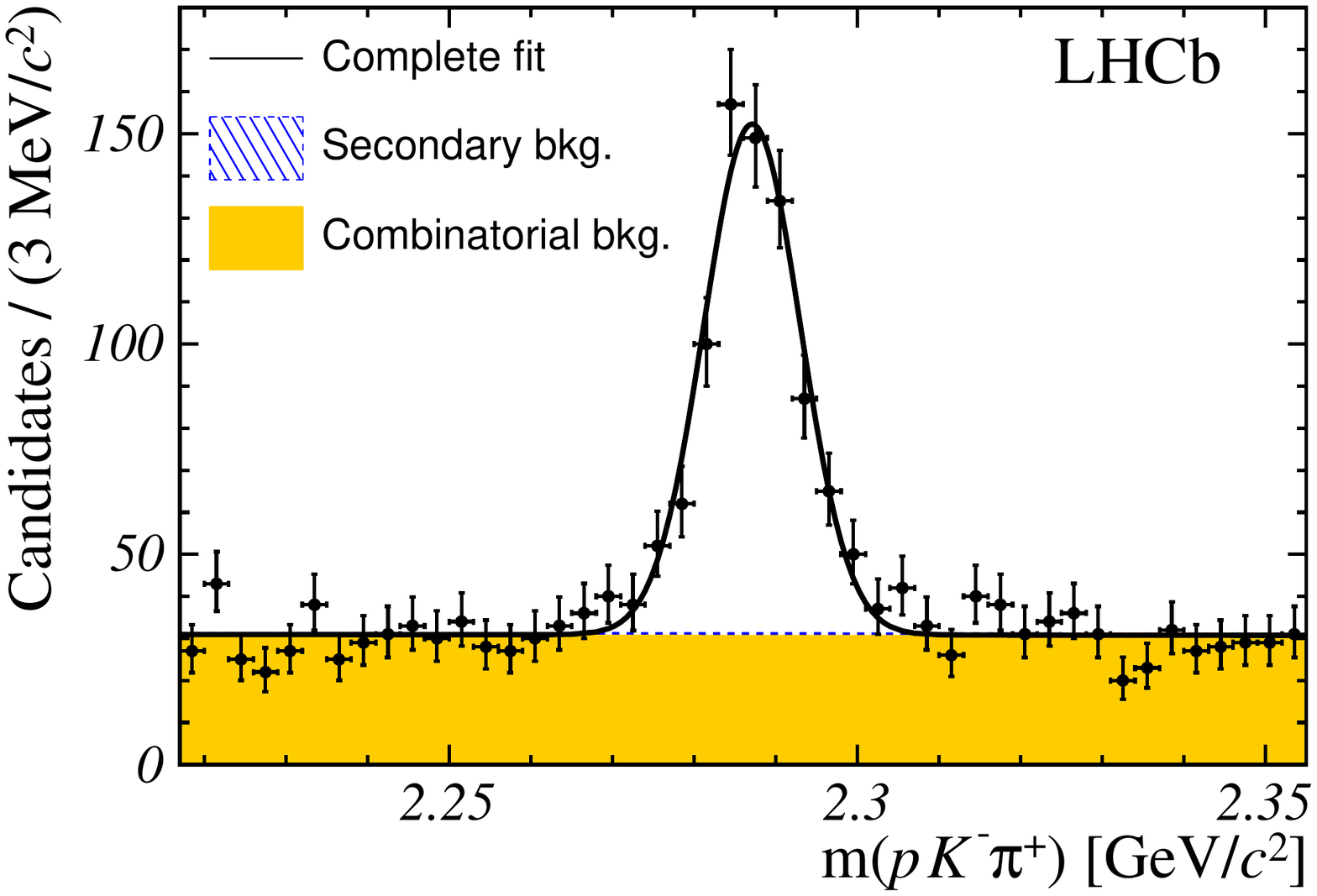}%
        \makebox[0cm][r]{\raisebox{0.19\textheight}[0cm]{\protect\subref{fig:Lcp:m:m}}\hspace{0.04\textwidth}}
      }%
      \subfloat{\label{fig:Lcp:m:ip}%
        \includegraphics[width=0.495\textwidth]{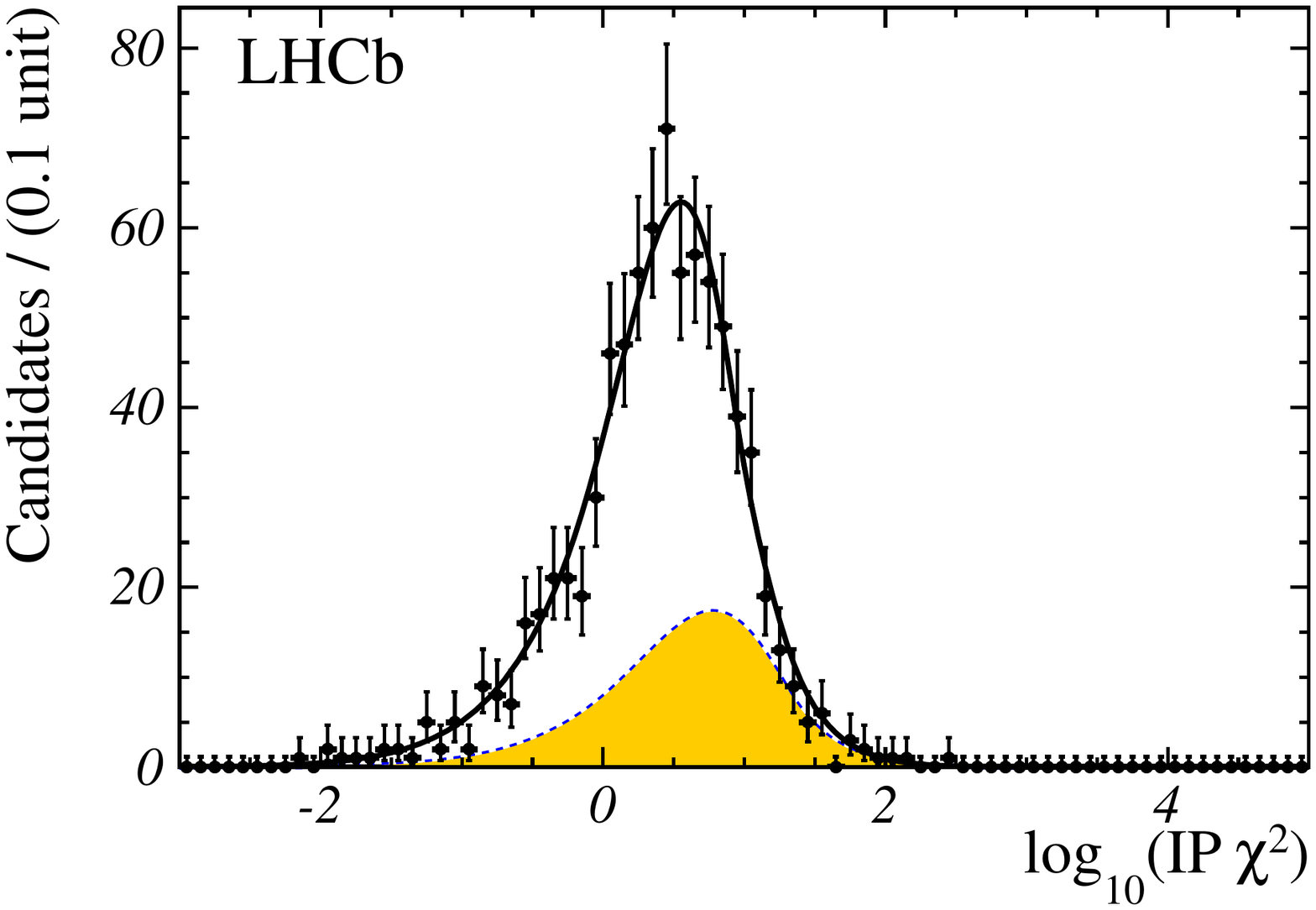}%
        \makebox[0cm][r]{\raisebox{0.19\textheight}[0cm]{\protect\subref{fig:Lcp:m:ip}}\hspace{0.37\textwidth}}
      }

      \caption[\DspTophipip and \LambdacpTopKmpip mass and \logipchisq distributions]{
        \small Mass and \logipchisq distributions for selected \DspTophipip
        and \LambdacpTopKmpip candidates showing
        \protect\subref{fig:Ds:m:m} the masses of the \Dsp candidates,
        \protect\subref{fig:Ds:m:ip} the \logipchisq distribution of
        \Dsp candidates for a mass window of \mbox{$\pm 8\MeVcc$}
        (approximately \mbox{$\pm 2\sigma$}) around the fitted
        $m(\Pphi(\Km\Kp)\pip)$ peak,
        \protect\subref{fig:Lcp:m:m} the masses of the \Lcp candidates,
        and \protect\subref{fig:Lcp:m:ip} the \logipchisq distribution
        of \Lcp candidates for a mass window of \mbox{$\pm 12\MeVcc$}
        (approximately \mbox{$\pm 2\sigma$}) around the fitted
        $m(\proton\Km\pip)$ peak.
        Projections of likelihood fits to the full data samples are shown
        with components as indicated in the legends.
      \label{fig:Ds:m}
      \label{fig:Lcp:m}}

    \end{figure}

  \afterpage{\clearpage}

  We factorise the efficiencies for reconstructing and selecting signal decays
  into components that are measured with independent studies.
  The particle identification (PID) efficiencies for pions, kaons, and protons
  are measured in data in
  bins of track $\pT$ and pseudorapidity, $\eta$, using high purity samples of
  pions, kaons, and protons from \KS, $\Pphi(1020)$, and
  \Lambdares decays.
  The effective total PID efficiency for each $(\pT, y)$ bin of each charmed
  hadron decay mode is determined by calculating the average efficiency over
  the bin using these final state PID efficiencies and the final state
  $(\pT, \eta)$ distributions from simulated decays.
  The efficiencies of the remaining selection criteria are determined from
  studies with the full event simulation.

%% Yield determination
%% ------------------------------------------------------------ %%
\subsection{Determination of signal yields}
\label{sec:analysis:yld}

  We use multidimensional extended maximum likelihood fits to the mass
  and \logipchisq distributions to determine the prompt signal yields.
  For the \DstarpTopipDz mode the \logipchisq of the daughter \Dz is used.
  The selected candidates contain \textit{secondary backgrounds}
  from signal decays produced in decays of \mbox{\bquark-hadrons}
  and \textit{combinatorial backgrounds}.
  The \DstarpTopipDz decay has two additional sources of background from
  \Dz decays combined with unrelated slow pion candidates:
  \textit{prompt random slow pion backgrounds} in which the \Dz mesons are
  produced at the PV and \textit {secondary random slow pion backgrounds} in
  which the \Dz mesons are produced in decays of \mbox{\bquark-hadrons}.
  The combinatorial backgrounds are separated from the remaining components
  with the reconstructed \Dz, \Dp, \Dsp, and \Lcp mass distributions.
  Analysis of the \logipchisq distributions allow separation of the
  prompt signal and secondary backgrounds.
  The additional random slow pion backgrounds in the \DstarpTopipDzToKmpip
  mode are identified in the distribution of the difference \Delm between the
  masses of the \Dstarp and \Dz candidates.
  Thus the prompt signal yields for \Dz, \Dp, \Dsp, and \Lcp decays are
  measured with two-dimensional fits to the mass and \logipchisq, and
  the prompt signal yields for \Dstarp decays are determined with
  three-dimensional fits to the \Dz candidate mass, \Delm, and \logipchisq.

  The extended likelihood functions are constructed from multidimensional 
  probability density functions (PDFs).
  For each class of events, the multidimensional PDF is the
  product of an appropriate one-dimensional PDF in each variable:
  \begin{description}
    \item[Prompt signal:]  The mass distributions are represented by Crystal
          Ball functions~\cite{Skwarnicki:1986xj}
          for \Dz decays (both direct and from \Dstarp mesons), double Gaussian
          functions for the \Dp and \Dsp modes, and a single Gaussian function
          for the \Lcp mode.
          The \Delm distribution for the \Dstarp mode is represented by a
          Crystal Ball function.
          The \logipchisq distributions are represented by bifurcated
          Gaussian functions with exponential tails defined as
  \begin{equation}
    \label{eq:yield:fit:agaussexp}
    \AGaussExp\mathopen{}\left(x; \mu, \sigma, \varepsilon, \rho_L, \rho_R\right)\mathclose{} =
      \begin{cases}
        \exp\mathopen{}\left(\frac{\rho_L^2}{2} + \frac{x - \mu}{\sigma \cdot (1 - \varepsilon)} \cdot \rho_L\right)\mathclose{}
          & \text{if } x < \mu - \rho_L \cdot \sigma \cdot (1 - \varepsilon), \\
        \exp\mathopen{}\left(-\frac{(x - \mu)^2}{2 \cdot \sigma^2 \cdot (1 - \varepsilon)^2}\right)\mathclose{}
          & \text{if } \mu - \rho_L \cdot \sigma \cdot (1 - \varepsilon) < x < \mu, \\
        \exp\mathopen{}\left(-\frac{(x - \mu)^2}{2 \cdot \sigma^2 \cdot (1 + \varepsilon)^2}\right)\mathclose{}
          & \text{if } \mu < x < \mu + \rho_R \cdot \sigma \cdot (1 + \varepsilon), \\
        \exp\mathopen{}\left(\frac{\rho_R^2}{2} - \frac{x - \mu}{\sigma \cdot (1 + \varepsilon)} \cdot \rho_R\right)\mathclose{}
          & \text{if } \mu + \rho_R \cdot \sigma \cdot (1 + \varepsilon) < x,
      \end{cases}
  \end{equation}
          where $\mu$ is the mode of the distribution,
          $\sigma$ is the average of the left and right Gaussian widths,
          $\varepsilon$ is the asymmetry of the left and right Gaussian widths,
          and $\rho_{L(R)}$ is the exponential coefficient for the left (right)
          tail.

    \item[Secondary backgrounds:]
          The functions representing the mass (and \Delm) distributions are
          identical to those used for the prompt signal in each case.
          The \logipchisq distributions are represented by \AGaussExp functions.

    \item[Combinatorial backgrounds:]
          The mass distributions are represented by first order polynomials.
          The \logipchisq distributions are represented by \AGaussExp functions.
          The \Delm distribution for the \Dstarp mode is represented by a
          power-law function $C \left(\Delm - M_{\pion}\right)^{p}$
          where the exponent $p$ is a free parameter;
          $M_{\pion}$ is the pion mass and $C$ is a normalisation
          constant.

    \item[Prompt random slow pion backgrounds] (\Dstarp only):
          The functions representing the mass and \logipchisq distributions
          are identical to those used for the prompt signal.
          The function representing the \Delm distribution is the same power
          law function as that used for the combinatorial backgrounds.

    \item[Secondary random slow pion backgrounds] (\Dstarp only):
          The functions representing the mass and \logipchisq distributions
          are identical to those used for the secondary backgrounds.
          The function representing the \Delm distribution is the same power
          law function as that used for the combinatorial backgrounds.
  \end{description}
  Shape parameters for the \logipchisq distributions of combinatorial
  backgrounds are fixed based on fits to the mass sidebands.
  Those of the prompt signal, secondary backgrounds, and random slow pion
  backgrounds are fixed based on fits to simulated events.
  \Figsref{fig:Dz:m}{fig:Lcp:m} show the results of single fits to
  the full \ptrange, \yrange kinematic region.

  The extended maximum likelihood fits are performed for each \pT-$y$ bin.
  We simultaneously fit groups of adjacent bins constraining
  to the same value several parameters that are expected to vary slowly across
  the kinematic region.
  The secondary background component in the \Lcp mode is too small to be
  measured reliably.
  We set its yield to zero when performing the fits and adopt
  a systematic uncertainty of 3\% to account
  for the small potential contamination from secondary production.

%% Systematics
%% ------------------------------------------------------------ %%
\subsection{Systematic uncertainties}
\label{sec:analysis:syst}

  There are three classes of systematic uncertainties: globally correlated
  sources, sources that are correlated between bins but uncorrelated between
  decay modes, and sources that are uncorrelated between bins and decay modes.
  The globally correlated contributions are the uncertainty on the
  measured luminosity and the uncertainty on the tracking efficiency.
  The former is a uniform $3.5\%$ for each mode.
  The latter is $3\%$ per final state track in the \Dz, \Dp, \Dsp, and \Lcp
  measurements and $4\%$ for the slow pion in the \Dstarp measurement.
  We adopt the uncertainty of the branching fractions as a
  bin-correlated systematic uncertainty.
  Systematic uncertainties of the reconstruction and selection efficiencies
  include contributions from the limited size of the
  simulated samples, failures in the association between generated and
  reconstructed particles in the simulation, differences between the observed
  and simulated distributions of selection variables, and differences between
  the simulated and actual resonance models in the \Dp and \Lcp measurements.
  The yield determination includes uncertainties from the fit models,
  from peaking backgrounds due to mis-reconstructed charm cross-feed,
  and from potential variations in the yields of secondary backgrounds.
  Where possible, the sizes of the systematic uncertainties are evaluated
  independently for each bin.
  The sources of systematic uncertainties are uncorrelated, and the total
  systematic uncertainty in each bin of each mode is determined by adding
  the systematic uncertainties in quadrature.
  \Tabref{tab:all_sys} summarises the systematic uncertainties.

\begin{table}[tp]
\caption[Overview of systematic uncertainties]{\label{tab:all_sys}
  \small Overview of systematic uncertainties and their values, expressed as
  relative fractions of the \xsectxtadj measurements in percent (\%).
  Uncertainties that are computed bin-by-bin are expressed as ranges giving
  the minimum to maximum values of the bin uncertainties.
  The correlated and uncorrelated uncertainties are shown as discussed in the
  text.
}
  %% $Id: $

\begin{tabular*}{\linewidth}{@{\extracolsep{\fill}}lccccc} \hline

Source  
                      & \Dz
                      & \Dstarp
                      & \Dp
                      & \Dsp
                      & \Lcp \\ \hline

%% Selection efficiency uncertainties
Selection and reconstruction (correlated)
                      & 1.6   %% \Dz
                      & 2.6   %% \Dstarp
                      & 4.3   %% \Dp Kpipi
                      & 5.3   %% \Dsp
                      & 0.4 \\ %% \Lcp 

\phantom{Selection and reconstruction} (uncorrelated)
                      & 1--12   %% \Dz
                      & 3--9    %% \Dstarp
                      & 1--10   %% \Dp Kpipi
                      & 4--9    %% \Dsp
                      & 5--17 \\ %% \Lcp 

%% Yield extraction
Yield determination (correlated)
                      & 2.5   %% \Dz
                      & 2.5   %% \Dstarp
                      & 0.5   %% \Dp Kpipi
                      & 1.0   %% \Dsp
                      & 3.0 \\ %% \Lcp 

\phantom{Yield determination} (uncorrelated)
                      & --   %% \Dz
                      & --   %% \Dstarp
                      & 1--5  %% \Dp Kpipi
                      & 2--14 %% \Dsp
                      & 4--9 \\ %% \Lcp 

%% PID
PID efficiency
                      & 1--5   %% \Dz
                      & 1--5   %% \Dstarp
                      & 6--19  %% \Dp Kpipi
                      & 1--15  %% \Dsp
                      & 5--9 \\ %% \Lcp 

%% Correlated stuff
Tracking efficiency
                      &  6 %% \Dz
                      & 10 %% \Dstarp
                      &  9 %% \Dp Kpipi
                      &  9 %% \Dsp
                      &  9 \\ %% \Lcp 

Branching fraction
                      &  1.3 % 0.05 / 3.89   %% \Dz
                      &  1.5 % 0.04 / 2.65   %% \Dstarp
                      &  2.1 % 0.19 / 9.13   %% \Dp Kpipi
                      &  5.8 % 0.13 / 2.24   %% \Dsp
                      & 26.0 \\ % 1.3  / 5.0 \\ %% \Lcp

Luminosity
                      & 3.5     %% \Dz
                      & 3.5     %% \Dstarp
                      & 3.5     %% \Dp
                      & 3.5     %% \Dsp
                      & 3.5 \\  %% \Lcp

\hline

\end{tabular*}

\end{table}

  As cross-checks, additional \xsectxtadj measurements are performed with the
  decay modes \DzToKmpippimpip and \DpTophipip and with a
  selection of \DzToKmpip decays that does not use particle identification
  information.
  Their results are in agreement with the results from our nominal
  measurements.

%%% Local Variables: 
%%% mode: latex
%%% TeX-master: t
%%% End: 
       % \label{sec:analysis}

\section{\Xsectxtnoun measurements}
\label{sec:results}

The signal yields determined from the data allow us to measure the
differential \xsecstxtnoun as functions of \pT and $y$
in the range \ptrange and \yrange.
The differential \xsectxtnoun for producing hadron species $\PH_c$ or its
charge conjugate in bin $i$, \dxdyinline{\ensuremath{\sigma_i(\PH_c)}}{\pT},
integrated over the $y$ range of the bin
is calculated with the relation 
\begin{linenomath}
  \begin{equation}
    \altdxdy{\sigma_i(\PH_c)}{\pT} = \frac{1}{\Delta\pT} \cdot \frac{N_i(\decay{\PH_c}{f} + \mbox{c.c.})}{\varepsilon_{i,\mathrm{tot}}(\decay{\PH_c}{f}) \cdot \BR(\decay{\PH_c}{f}) \cdot \lum_{\mathrm{int}}},
    \label{eq:xsec}
  \end{equation}
\end{linenomath}
where $\Delta\pT$ is the width in \pT of bin $i$, typically $1\GeVc$,
$N_i(\decay{\PH_c}{f} + \mbox{c.c.})$ is the measured yield of $\PH_c$
and their charge conjugate decays
in bin $i$, $\BR(\decay{\PH_c}{f})$ is the branching fraction of the
decay, $\varepsilon_{i,\mathrm{tot}}(\decay{\PH_c}{f})$ is the total efficiency
for observing the signal decay in bin $i$, and
$\lum_{\mathrm{int}} = \lumipmerr$ is the integrated luminosity of the sample.
The following branching fractions from \refref{PDG2012} are used:
\mbox{$\BF(\DpToKmpippip) = (9.13 \pm 0.19)\%$},
\mbox{$\BF(\DstarpTopipDzToKmpip) = (2.63 \pm 0.04)\%$},
\mbox{$\BF(\LambdacpTopKmpip) = (5.0 \pm 1.3)\%$},
and \mbox{$\BF(\DzDzbToKmpip) = (3.89 \pm 0.05)\%$},
where the last is the sum of Cabibbo-favoured and doubly Cabibbo-suppressed
branching fractions.
For the \Ds measurement we use the branching fraction of \DspToKmKppip in a
$\pm 20\,\MeVc$ window around the $\phi(1020)$ mass:
\mbox{$\BF(\DspTophipip) = (2.24 \pm 0.13)\%$}~\cite{Alexander:2008cqa}.
The measured differential \xsecstxtnoun are tabulated in the appendix.
Bins with a sample size insufficient to produce a
measurement with a total relative uncertainty of less than $50\%$ are
discarded.

Theoretical expectations for the production \xsecstxtnoun of charmed
hadrons have been calculated by Kniehl \etal using the GMVFNS
scheme~\cite{Kniehl:2004fy,Kniehl:2005gz,Kniehl:2005ej,Kneesch:2007ey,Kniehl:2009ar,Kniehl:2012ti}
and Cacciari \etal, using the FONLL
approach~\cite{Cacciari:1998it,Cacciari:2003zu,Cacciari:2005uk,Cacciari:2012ny}.
Both groups have provided differential \xsecstxtnoun as functions
of \pT and integrated over bins in $y$.

\begin{figure}[p]
  \centering
    \subfloat{\label{fig:lhcb:Dz:a}%
      \includegraphics[width=0.495\textwidth]{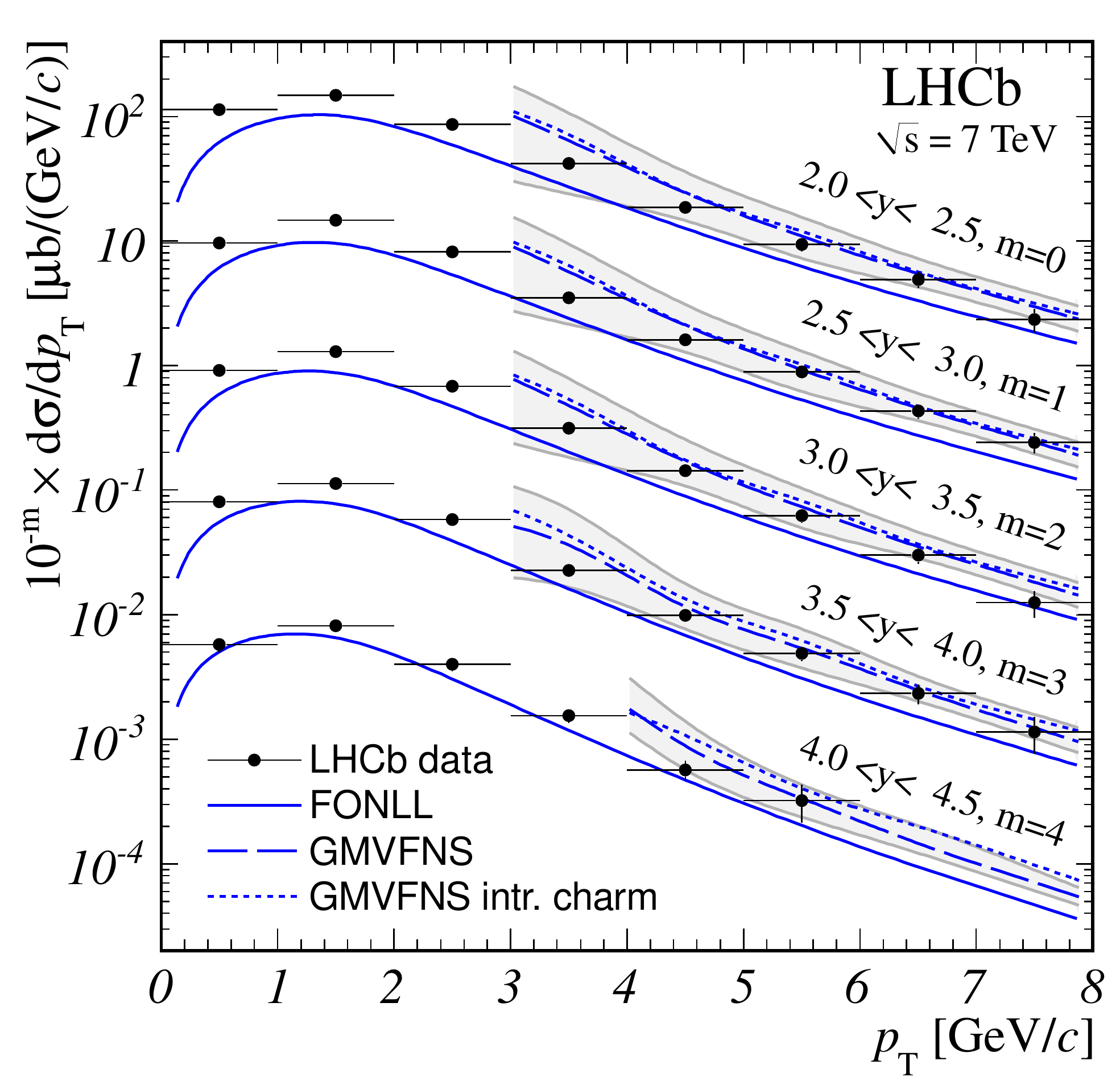}%
      \makebox[0cm][r]{\raisebox{0.31\textheight}[0cm]{\protect\subref{fig:lhcb:Dz:a}}\hspace{0.14\textwidth}}
    }%
    \subfloat{\label{fig:lhcb:Dp:b}%
      \includegraphics[width=0.495\textwidth]{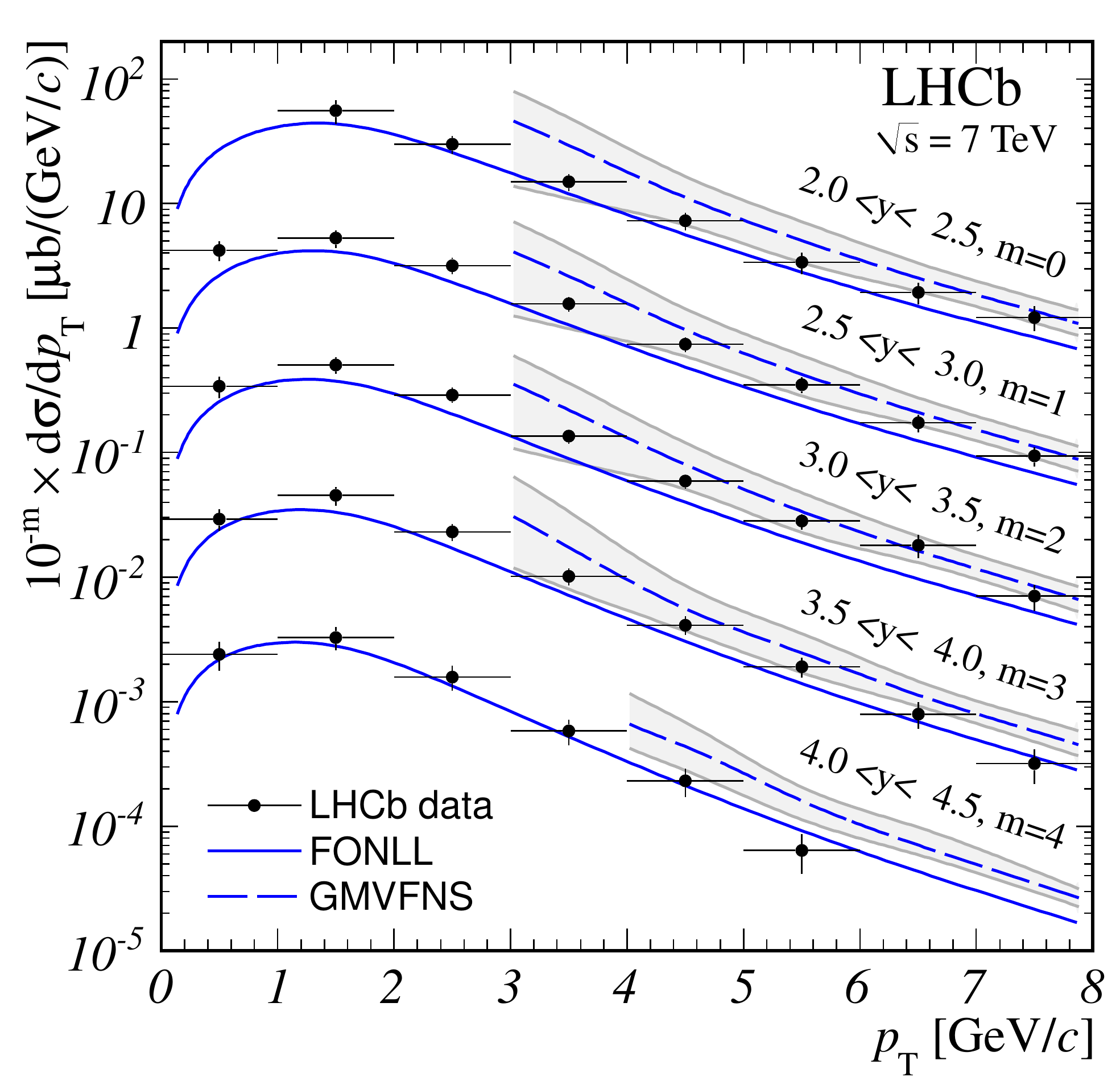}%
      \makebox[0cm][r]{\raisebox{0.31\textheight}[0cm]{\protect\subref{fig:lhcb:Dp:b}}\hspace{0.14\textwidth}}
    }

    \subfloat{\label{fig:lhcb:dstar:a}%
      \includegraphics[width=0.495\textwidth]{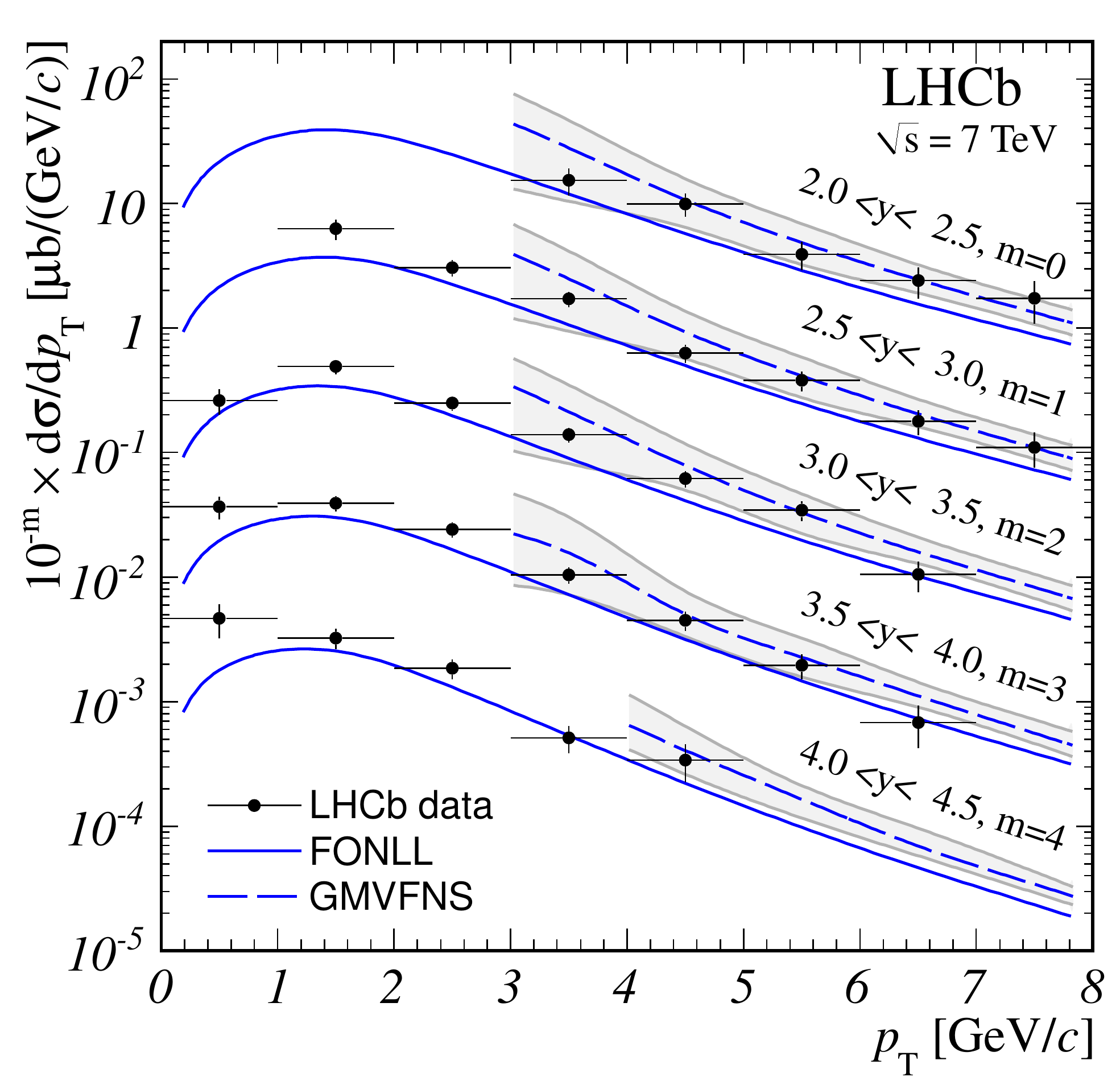}%
      \makebox[0cm][r]{\raisebox{0.31\textheight}[0cm]{\protect\subref{fig:lhcb:dstar:a}}\hspace{0.14\textwidth}}
    }%
    \subfloat{\label{fig:lhcb:Dsp:b}%
      \includegraphics[width=0.495\textwidth]{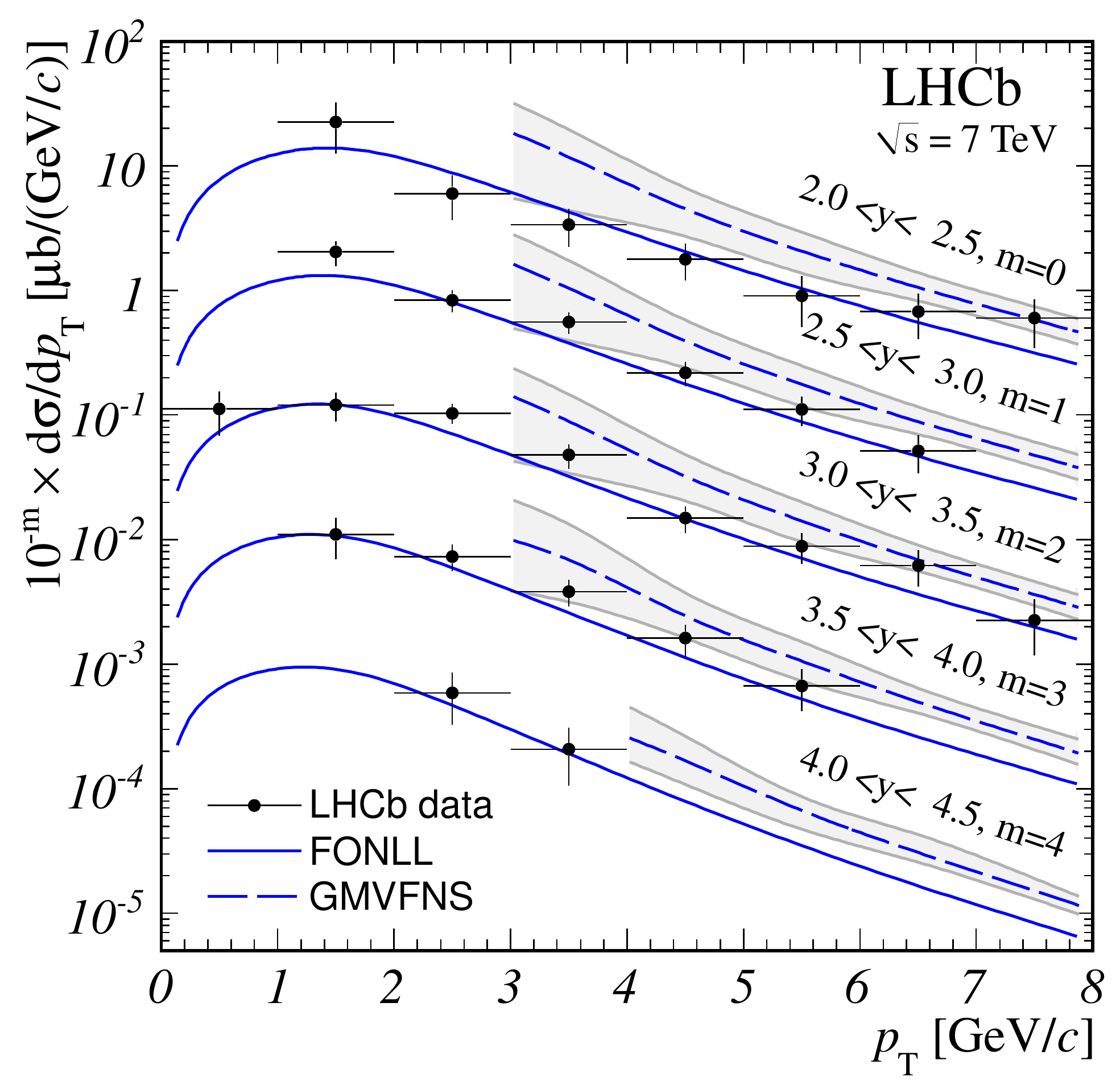}
      \makebox[0cm][r]{\raisebox{0.31\textheight}[0cm]{\protect\subref{fig:lhcb:Dsp:b}}\hspace{0.14\textwidth}}
    }

   \caption[\PD meson \xsecstxtnoun]{
     \small Differential \xsecstxtnoun for \protect\subref{fig:lhcb:Dz:a} \Dz,
     \protect\subref{fig:lhcb:Dp:b} \Dp,
     \protect\subref{fig:lhcb:dstar:a} \Dstarp,
     and \protect\subref{fig:lhcb:Dsp:b} \Dsp
     meson production compared to theoretical predictions.
     The \xsecstxtnoun for different $y$ regions are shown as functions of
     \pt.
     The $y$ ranges are shown as separate curves and associated sets of
     points scaled by factors $10^{-m}$, where the exponent $m$ is shown on
     the plot with the $y$ range.
     The error bars associated with the data points show the sum in quadrature
     of the statistical and total systematic uncertainty.
     The shaded regions show the
     range of theoretical uncertainties for the GMVFNS prediction.
     \label{fig:lhcb:Dz}
     \label{fig:lhcb:Dp}
     \label{fig:lhcb:dstar}
     \label{fig:lhcb:Dsp}}
  \end{figure}

The FONLL calculations use the
CTEQ~6.6~\cite{Nadolsky:2008zw} parameterisation of the parton densities.
They include estimates of theoretical uncertainties due to the charm quark
mass and the renormalisation and factorisation scales.
However, we display only the central values in
\figsref{fig:lhcb:Dz}{fig:results:lambdacp}.
The theoretical calculations assume unit transition probabilities from a
primary charm quark to the exclusive hadron state.
The actual transition probabilities that we use to convert the predictions to
measurable \xsecstxtnoun are those quoted by
\refref{Amsler:2008zzb:Frag}, based on measurements from $\ep\en$\/
colliders close to the $\Upsilon(4S)$ resonance:
\mbox{$f(\cquarkto{\Dz}) = 0.565 \pm 0.032$},
\mbox{$f(\cquarkto{\Dp}) = 0.246 \pm 0.020$},
\mbox{$f(\cquarkto{\Dstarp}) = 0.224 \pm 0.028$},
\mbox{$f(\cquarkto{\Dsp}) = 0.080 \pm 0.017$},
and \mbox{$f(\cquarkto{\Lcp}) = 0.094 \pm 0.035$}.
Note that the transition probabilities do not sum up to unity,
since, \eg, $f(\cquarkto{\Dz})$ has an overlapping contribution
from $f(\cquarkto{\Dstarp})$.
No dedicated calculation for \Dsp production is available.
The respective prediction was obtained by scaling the kinematically similar
\Dstarp prediction by the ratio
\mbox{$f(\cquarkto{\Dsp})/f(\cquarkto{\Dstarp})$}.

The GMVFNS calculations include theoretical
predictions for all hadrons studied in our analysis.
Results were provided for \mbox{$\pT > 3\GeVc$}.
The uncertainties from scale variations were determined only for the
case of \Dz production.
The relative sizes of the uncertainties for the other hadron species are
assumed to be the same as those for the \Dz.
Here the CTEQ~6.5~\cite{Tung:2006tb} set of parton densities was used.
Predictions for \Dz mesons were also provided using the
CTEQ~6.5c2~\cite{Pumplin:2007wg} parton densities with intrinsic charm.
As shown in \figref{fig:lhcb:Dz:a}, in the phase space region
of the present measurement the effect of intrinsic charm is predicted to be
small.
The GMVFNS theoretical framework includes the convolution with fragmentation
functions describing the transition \cquarkto{\ensuremath{\PH_c}}
that are normalised to the respective total transition
probabilities~\cite{Kneesch:2007ey}.
The fragmentation functions are results of a fit to production
measurements at \epem colliders, where no attempt was made in the fit to
separate direct production and feed-down from higher resonances.

To compare the theoretical calculations to our measurements,
the theoretical differential \xsecstxtnoun were integrated over
the \pT bins and then divided by the bin width $\Delta\pT$.
The integration was performed numerically with a third-order spline
interpolation of the differential \xsecstxtnoun.

The measured \xsecstxtnoun compared to the theoretical predictions are
shown in \figsref{fig:lhcb:Dz}{fig:results:lambdacp}.
For better visibility, theoretical predictions are displayed as smooth
curves such that the value at the bin centre corresponds to the
differential \xsectxtnoun calculated in that bin.
The data points with their uncertainties, which are always drawn at the bin
centre, thus can be directly compared with theory.
The predictions agree well with our measurements, generally bracketing the
observed values between the FONLL and GMVFNS calculations.

  \begin{figure}[p]
  \centering
    \includegraphics[width=0.495\textwidth]{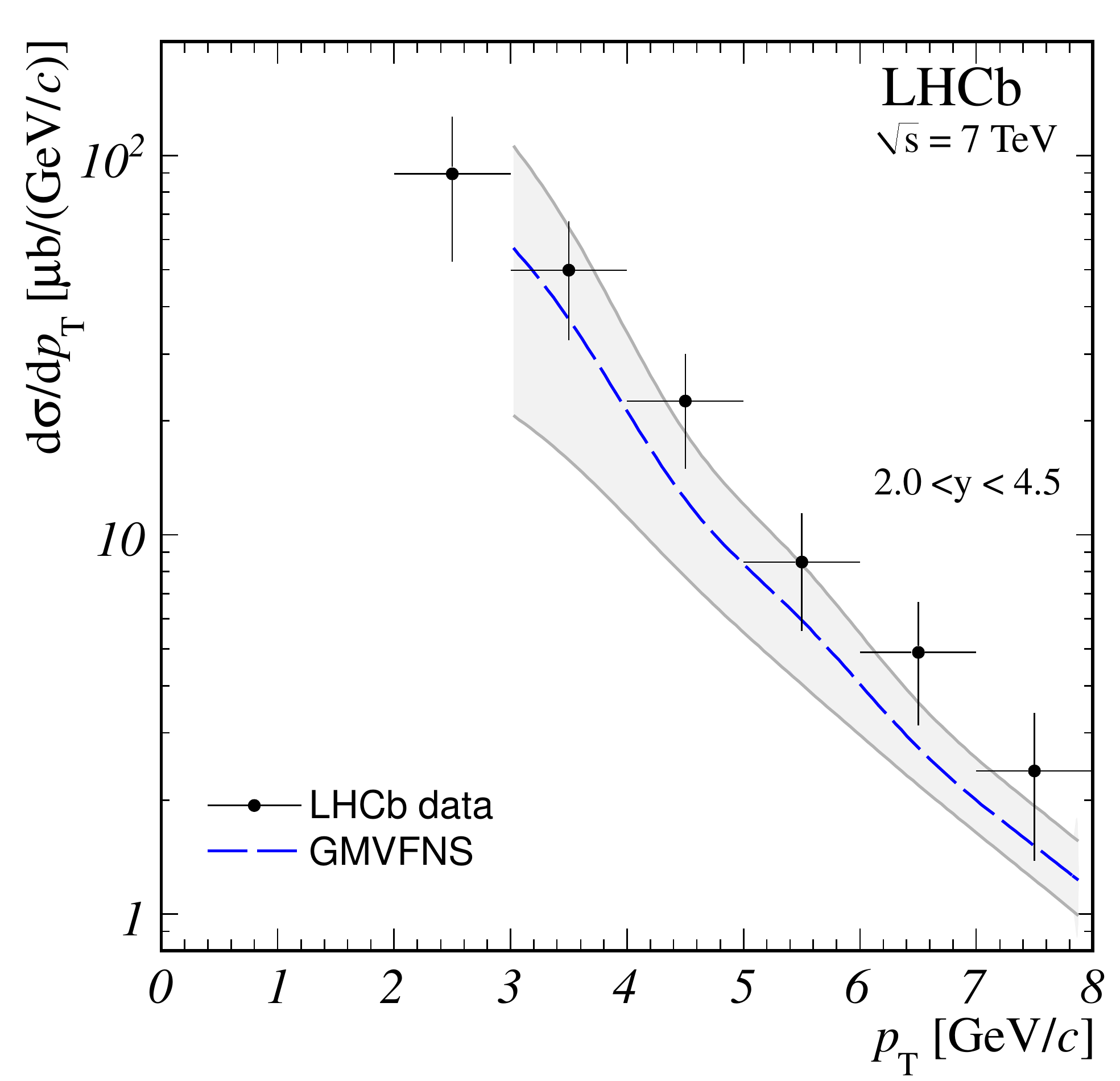}

   \caption[\Lc \xsecstxtnoun]{
     \small Differential \xsecstxtnoun for \Lcp baryon production compared to
     the theoretical prediction from the GMVFNS scheme.
     The error bars associated with the data points show the sum in quadrature
     of the statistical and total systematic uncertainty.
     The shaded region shows the
     range of theoretical uncertainty for the theoretical prediction.
     \label{fig:results:lambdacp}}
  \end{figure}

  \afterpage{\clearpage}

%%% Local Variables: 
%%% mode: latex
%%% TeX-master: t
%%% End: 
        % \label{sec:results}

\section{Production ratios and integrated \xsecstxtnoun}
\label{sec:xsecs}

  Charmed hadron production ratios and total \xsecstxtnoun
  are determined for the kinematic range \ptrange and \yrange.
  Bins where the relative uncertainty on the yield exceeds $50\%$ (left blank
  in \Tabsref{tab:xsec:LambdacpPKP-y}{tab:xsec:DsPhiPi} of the appendix)
  are not used.
  Instead, the \xsecstxtnoun are extrapolated from the remaining bins
  with predictions obtained from \pythiasix.
  The extrapolation factors are computed as the ratios of the predicted
  \xsecstxtnoun integrated over \ptrange and \yrange to the predicted
  \xsecstxtnoun integrated over the well measured bins for each of four tunes
  of \pythiasix: \LHCb-tune~\cite{LHCb-PROC-2010-056},
  Perugia~0, Perugia~NOCR, and Perugia~2010~\cite{Skands:2010ak}.
  The mean of these four ratios is used as a multiplicative factor to
  extrapolate the sum of the well measured bins to the full kinematic range
  under study.
  The root mean square of the four ratios is taken as a systematic uncertainty
  associated with the extrapolation.
  We confirm that this procedure gives uncertainties of appropriate size by
  examining the variance of the ratios for individual well measured bins.
  The resulting integrated \xsecstxtnoun for each hadron species are given
  in \Tabref{tab:results:total:input}.

  \begin{table}[tp]
  \caption[Integrated production \xsecstxtnoun into \LHCb acceptance]{
    \small Open charm production \xsecstxtnoun in the kinematic range
    \ptrange and \yrange.
    The computation of the extrapolation factors is described in the text.
    The first uncertainty is statistical, the second is systematic, and the
    third is the contribution from the extrapolation factor.
   \label{tab:results:total:input}}

    \centering
    %% $Id: $
%% ====================================================================== %%

    \begin{tabular}{lrd{4}@{\,$\pm$\,}d{4}r*{3}{r@{\,$\pm$\,}}r} \hline

      & \multicolumn{4}{c}{Extrapolation factor}
      & \multicolumn{4}{c}{\Xsectxtnoun (\mubNS)} \\
      \hline

      \Dz
      & & 1.003 & 0.001 &
      & 1661 & 16 & 128 & 2 \\

      \Dp
      & & 1.067  & 0.013  &
      & 645 & 11 & 72 &  8 \\

      \Dstarp
      & & 1.340  & 0.037  &
      & 677 & 26 & 77 & 19 \\

      \Dsp
      & & 1.330  & 0.056  &
      & 197 & 14 & 26 & 8 \\

      \Lcp
      & & 1.311  & 0.077  &
      & 233 & 26 & 71 & 14 \\

      \hline

    \end{tabular}

  \end{table}

  Accounting for the correlations among the sources of systematic uncertainty,
  we obtain the correlation matrix for the total uncertainties of the
  integrated \xsectxtadj measurements shown in
  \Tabref{tab:results:total:corr}.
  The ratios of the production \xsecstxtnoun in the kinematic range \ptrange
  and \yrange are given in \Tabref{tab:results:total:comb5}.

  \begin{table}[tp]
  \caption[Correlation matrix of measured integrated production \xsecstxtnoun]{
    \small Correlation matrix of the uncertainties of the integrated open
    charm production \xsecstxtnoun
    in the kinematic range \ptrange and \yrange.
    The first column restates measured values of the integrated
    \xsecstxtnoun.
   \label{tab:results:total:corr}}

    \centering
    %% $Id: $
%% ====================================================================== %%

  \begin{tabular}{l@{\,=\,}r@{\,\ensuremath{\pm}\,}r@{\protect\mub}*{4}{d{2}}} \hline

    \multicolumn{3}{c}{} 
    & \multicolumn{1}{c}{\rule[-1.5ex]{0pt}{4ex}\ensuremath{\sigma(\Dz)}} 
    & \multicolumn{1}{c}{\ensuremath{\sigma(\Dp)}}
    & \multicolumn{1}{c}{\ensuremath{\sigma(\Dstarp)}}
    & \multicolumn{1}{c}{\ensuremath{\sigma(\Dsp)}} \\ \hline

    $\sigma(\Dz)$     & 1661 & 129 &      &      &       &      \\
    $\sigma(\Dp)$     &  645 &  74 & 0.76 &      &       &      \\
    $\sigma(\Dstarp)$ &  677 &  83 & 0.77 & 0.73 &       &      \\
    $\sigma(\Dsp)$    &  197 &  31 & 0.55 & 0.52 &  0.53 &      \\
    $\sigma(\Lcp)$    &  233 &  77 & 0.26 & 0.25 &  0.25 & 0.18 \\
    \hline
  \end{tabular}

  \end{table}

  \begin{table}[tp]
  \caption[Ratios of integrated production \xsectxtnoun for analysed species]{
    \small \Xsectxtadj ratios for open charm production in the kinematic range
    \ptrange and \yrange.
    The numbers in the table are the ratios of the respective row/column.
   \label{tab:results:total:comb5}}

    \centering
    %% $Id: $
%% ====================================================================== %%

    \begin{tabular}{@{\extracolsep{\fill}}lcccc} \hline

      & \rule[-1.5ex]{0pt}{4ex}$\sigma(\Dz)$
      & $\sigma(\Dp)$
      & $\sigma(\Dstarp)$
      & $\sigma(\Dsp)$ \\
      \hline  

      $\sigma(\Dp)$
      & $0.389\pm 0.029$        % $\sigma(\Dz)$
      &                         % $\sigma(\Dp)$
      &
      &
      \\

      $\sigma(\Dstarp)$
      & $0.407\pm 0.033$        % $\sigma(\Dz)$
      & $1.049\pm 0.092$        % $\sigma(\Dp)$
      &                         % $\sigma(\Dstarp)$
      &
      \\

      $\sigma(\Dsp)$
      & $0.119\pm 0.016$        % $\sigma(\Dz)$
      & $0.305\pm 0.042$        % $\sigma(\Dp)$
      & $0.291\pm 0.041$        % $\sigma(\Dstarp)$
      &                         % $\sigma(\Dsp)$
      \\

      $\sigma(\Lcp)$
      & $0.140\pm 0.045$        % $\sigma(\Dz)$
      & $0.361\pm 0.116$        % $\sigma(\Dp)$
      & $0.344\pm 0.111$        % $\sigma(\Dstarp)$
      & $1.183\pm 0.402$        % $\sigma(\Dsp)$
      \\

      \hline
    \end{tabular}

  \end{table}

  Finally, we determine the total charm \xsectxtnoun contributing to charmed
  hadron production inside the acceptance of this study, \ptrange
  and \yrange.
  Combining our measurements $\sigma(\PH_c)$ with the corresponding
  fragmentation functions $f(\cquarkto{\PH_c})$ from
  \refref{Amsler:2008zzb:Frag} gives five estimates of
  $\sigma(\ccbar) = \sigma(\PH_c) / (2 f(\cquarkto{\PH_c}))$.
  The factor of $2$ appears in the denominator because we have defined
  $\sigma(\PH_c)$ to be the \xsectxtnoun to produce either $\PH_c$
  or its charge conjugate.
  A combination of all five measurements taking correlations into account gives
  \begin{equation*}
      \sigma(\ccbar)_{p_{\mathrm{T}} < 8\mathrm{\,Ge\kern -0.1em V\!/}c, \,2.0 < y < 4.5}
        =  \totccbarxsec,
  \end{equation*}
  The final uncertainty is that due to the fragmentation functions.

      % \label{sec:xsecs}

\section{Summary}
\label{sec:summary}

A measurement of charm production in $\proton\proton$ collisions at a
centre-of-mass energy of 7\TeV has been performed with the
\LHCb detector, based on an integrated luminosity of
\mbox{$\lum_{\mathrm{int}} = \lumitwosigfig$}.
\Xsectxtadj measurements with total uncertainties below 20\% have been
achieved.
The shape and absolute normalisation of the differential \xsecstxt for
\Dz/\Dzb, \Dpm, \Dstarpm, \Dspm, and \Lcpm hadrons are found to be in
agreement with theoretical predictions.
The ratios of the production \xsecstxtnoun for the five species under study
have been measured.
The \ccbar \xsectxtnoun for producing a charmed hadron
in the range \ptrange and \yrange is found to be
\mbox{\totccbarxsec}.

%%% Local Variables: 
%%% mode: latex
%%% TeX-master: t
%%% End: 
        % \label{sec:summary}

% Do not include this in analysis note and conference reports
\section*{Acknowledgements}

%% $Id: $

\noindent The authors are grateful to H.~Spiesberger, B.~A.~Kniehl, G.~Kramer,
and I.~Schienbein for providing theoretical \xsectxtadj predictions from
the \GMVFNStext (GMVFNS).
We thank M.~Mangano, M.~Cacciari, S.~Frixione, M.~Nason, and G.~Ridolfi
for supplying theoretical \xsectxtadj predictions using the
\FONLLtext (FONLL) approach.

We express our gratitude to our colleagues in the CERN
accelerator departments for the excellent performance of the LHC. We
thank the technical and administrative staff at the LHCb
institutes. We acknowledge support from CERN and from the national
agencies: CAPES, CNPq, FAPERJ and FINEP (Brazil); NSFC (China);
CNRS/IN2P3 and Region Auvergne (France); BMBF, DFG, HGF and MPG
(Germany); SFI (Ireland); INFN (Italy); FOM and NWO (The Netherlands);
SCSR (Poland); ANCS/IFA (Romania); MinES, Rosatom, RFBR and NRC
``Kurchatov Institute'' (Russia); MinECo, XuntaGal and GENCAT (Spain);
SNSF and SER (Switzerland); NAS Ukraine (Ukraine); STFC (United
Kingdom); NSF (USA). We also acknowledge the support received from the
ERC under FP7. The Tier1 computing centres are supported by IN2P3
(France), KIT and BMBF (Germany), INFN (Italy), NWO and SURF (The
Netherlands), PIC (Spain), GridPP (United Kingdom). We are thankful
for the computing resources put at our disposal by Yandex LLC
(Russia), as well as to the communities behind the multiple open
source software packages that we depend on.

% $Id: appendix.tex 30628 2013-01-28 11:08:12Z spradlin $
% ===============================================================================
% Purpose: appendix to the standard template: standard symbol alises from Ulrik
% Author: Tomasz Skwarnicki
% Created on: 2009-09-24
% ===============================================================================

\clearpage

{\noindent\bf\Large Appendix}

\appendix

\section*{Measured open charm \xsecstxtnoun}
\label{sec:app:xsec}

\Tabref{tab:xsec:LambdacpPKP-y} shows the production \xsecstxtnoun for \Lcp
baryons integrated over \mbox{$2 < \pT < 8\GeVc$} and over the rapidity range
of the $y$ bins.
The differential production \xsectxtadj values (integrated over the $y$ range
of the respective bin) plotted in
\figsref{fig:lhcb:Dz}{fig:results:lambdacp} are given in
\Tabsref{tab:xsec:LambdacpPKP-pT}{tab:xsec:DsPhiPi}.

\begin{table}[hbp]

  \caption[ \Xsecstxtnoun of \LambdacpTopKmpip ] {
    \small Bin-integrated production \xsecstxtnoun in \mub for prompt
    \mbox{\Lcp $+$ c.c.} baryons
    in bins of $y$ integrated over the range \mbox{$2 < \pT < 8\GeVc$}.
    The first uncertainty is statistical, and the second is the total
    systematic.
   \label{tab:xsec:LambdacpPKP-y}}

  %% $Id: $

  \begin{tabular*}{\linewidth}{@{\extracolsep{\fill}}ccccc} \hline

    \pT & \multicolumn{4}{c}{$y$} \\

    (\GeVcNS) & $(2.0, 2.5)$ & $(2.5, 3.0)$ & $(3.0, 3.5)$ & $(3.5, 4.0)$ \\ \hline

    $(2, 8)$ & $   21.4 \pm     8.1 \pm     7.2$ & $   49.9 \pm    11.6 \pm    15.6$ & $   62.9 \pm     7.0 \pm    18.8$ & $   44.2 \pm     8.6 \pm    13.2$ \\

    \hline

  \end{tabular*}

\end{table}

\begin{table}[hbp]

  \caption[ \Xsecstxtnoun of \LambdacpTopKmpip ] {
    \small Differential production \xsecstxtnoun,
    \dxdy{\ensuremath{\sigma}}{\pT}, in $\mub/(\GeVcNS)$ for prompt
    \mbox{\Lcp $+$ c.c.} baryons
    in bins of $\pT$ integrated over the rapidity range \yrange.
    The first uncertainty is statistical, and the second is the total
    systematic.
   \label{tab:xsec:LambdacpPKP-pT}}

  \centering
    %% $Id: $

  \begin{tabular}{cd{1}@{\,\ensuremath{\pm}\,}d{1}@{\,\ensuremath{\pm}\,}d{1}} \hline

    \pT & \multicolumn{3}{c}{$y$} \\
    (\GeVcNS) & \multicolumn{3}{c}{$(2.0, 4.5)$} \\ \hline

    $(2, 3)$  &    89.6 &    17.8 &    32.6 \\
    $(3, 4)$  &    49.8 &     7.9 &    15.3 \\
    $(4, 5)$  &    22.5 &     3.1 &     6.9 \\
    $(5, 6)$  &     8.5 &     1.4 &     2.6 \\
    $(6, 7)$  &     4.9 &     0.9 &     1.5 \\
    $(7, 8)$  &     2.4 &     0.6 &     0.8 \\

    \hline

  \end{tabular}

\end{table}

\begin{sidewaystable}[p]

  \caption[ \Xsecstxtnoun of \DzToKmpip ] {
    \small Differential production \xsecstxtnoun,
    \dxdy{\ensuremath{\sigma}}{\pT}, in $\mub/(\GeVcNS)$ for prompt
    \mbox{\Dz $+$ c.c.} mesons in bins of $(\pT, y)$.
    The first uncertainty is statistical, and the second is the total
    systematic.
   \label{tab:xsec:DzeroKP}}

  %% $Id: $

  \begin{tabularx}{\linewidth}{C*{5}{d{2}@{\,$\pm$\,}d{2}@{\,$\pm$\,}d{2}}} \hline

    \pT & \multicolumn{15}{c}{$y$} \\

    (\GeVcNS)  & \multicolumn{3}{c}{$(2.0, 2.5)$} & \multicolumn{3}{c}{$(2.5, 3.0)$} & \multicolumn{3}{c}{$(3.0, 3.5)$} & \multicolumn{3}{c}{$(3.5, 4.0)$} & \multicolumn{3}{c}{$(4.0, 4.5)$}\\ \hline

    $(0, 1)$  &  113.58 &    5.45 &   10.45 &   96.51 &    3.49 &    8.10 &   90.99 &    3.67 &    7.24 &   80.41 &    4.19 &    6.30 &   57.37 &    5.37 &    5.10 \\
    $(1, 2)$  &  147.06 &    5.78 &   12.45 &  146.54 &    4.08 &   12.16 &  129.43 &    3.89 &   10.19 &  112.64 &    4.52 &    8.95 &   81.57 &    5.20 &    7.02 \\
    $(2, 3)$  &   85.95 &    3.18 &    6.80 &   82.07 &    2.10 &    6.58 &   68.48 &    1.90 &    5.40 &   58.25 &    2.02 &    4.70 &   39.87 &    2.56 &    3.78 \\
    $(3, 4)$  &   41.79 &    1.78 &    3.82 &   34.86 &    1.10 &    2.82 &   31.30 &    1.05 &    2.47 &   22.65 &    1.00 &    2.13 &   15.50 &    1.29 &    1.51 \\
    $(4, 5)$  &   18.61 &    0.98 &    1.73 &   16.11 &    0.67 &    1.49 &   14.36 &    0.66 &    1.15 &    9.89 &    0.62 &    0.94 &    5.69 &    0.87 &    0.60 \\
    $(5, 6)$  &    9.35 &    0.66 &    0.90 &    8.85 &    0.48 &    0.84 &    6.23 &    0.41 &    0.60 &    4.88 &    0.43 &    0.48 &    3.22 &    0.98 &    0.46 \\
    $(6, 7)$  &    4.92 &    0.51 &    0.49 &    4.31 &    0.38 &    0.43 &    2.99 &    0.33 &    0.30 &    2.33 &    0.34 &    0.25 & \multicolumn{3}{c}{}   \\
    $(7, 8)$  &    2.34 &    0.42 &    0.26 &    2.41 &    0.36 &    0.26 &    1.25 &    0.27 &    0.14 &    1.14 &    0.35 &    0.16 & \multicolumn{3}{c}{}   \\

    \hline

  \end{tabularx}

  \vspace{2ex}

  \caption[ \Xsecstxtnoun of \DpToKmpippip ] {
    \small Differential production \xsecstxtnoun,
    \dxdy{\ensuremath{\sigma}}{\pT}, in $\mub/(\GeVcNS)$ for prompt
    \mbox{\Dp $+$ c.c.} mesons in bins of $(\pT, y)$.
    The first uncertainty is statistical, and the second is the total
    systematic.
   \label{tab:xsec:DplusKPP}}

  %% $Id: $

  \begin{tabularx}{\linewidth}{C*{5}{d{2}@{\,$\pm$\,}d{2}@{\,$\pm$\,}d{2}}} \hline

    \pT & \multicolumn{15}{c}{$y$} \\

    (\GeVcNS)  & \multicolumn{3}{c}{$(2.0, 2.5)$} & \multicolumn{3}{c}{$(2.5, 3.0)$} & \multicolumn{3}{c}{$(3.0, 3.5)$} & \multicolumn{3}{c}{$(3.5, 4.0)$} & \multicolumn{3}{c}{$(4.0, 4.5)$}\\ \hline

    $(0, 1)$  & \multicolumn{3}{c}{} &   42.11 &    2.92 &    7.21 &   34.00 &    1.78 &    6.29 &   29.32 &    1.89 &    5.52 &   24.01 &    2.94 &    5.45 \\
    $(1, 2)$  &   \phantom{1}55.56 &    6.79 &    9.89 &   \phantom{11}52.72 &    2.27 &    8.31 &   \phantom{11}50.74 &    1.66 &    7.68 &   \phantom{11}45.26 &    1.70 &    7.56 &   32.87 &    2.47 &    6.59 \\
    $(2, 3)$  &   29.86 &    2.38 &    4.40 &   31.79 &    1.09 &    4.57 &   29.03 &    0.87 &    3.99 &   23.09 &    0.84 &    3.45 &   15.79 &    1.17 &    3.43 \\
    $(3, 4)$  &   14.97 &    1.04 &    2.14 &   15.69 &    0.57 &    2.10 &   13.53 &    0.48 &    1.71 &   10.15 &    0.45 &    1.49 &    5.84 &    0.55 &    1.25 \\
    $(4, 5)$  &    7.26 &    0.54 &    1.01 &    7.44 &    0.33 &    0.96 &    5.89 &    0.27 &    0.74 &    4.12 &    0.26 &    0.65 &    2.31 &    0.32 &    0.50 \\
    $(5, 6)$  &    3.37 &    0.31 &    0.58 &    3.51 &    0.21 &    0.46 &    2.81 &    0.18 &    0.36 &    1.90 &    0.16 &    0.31 &    0.64 &    0.18 &    0.14 \\
    $(6, 7)$  &    1.93 &    0.21 &    0.31 &    1.73 &    0.14 &    0.23 &    1.81 &    0.14 &    0.36 &    0.80 &    0.10 &    0.17 & \multicolumn{3}{c}{}   \\
    $(7, 8)$  &    1.22 &    0.17 &    0.22 &    0.94 &    0.11 &    0.13 &    0.70 &    0.09 &    0.14 &    0.32 &    0.07 &    0.07 & \multicolumn{3}{c}{}   \\

    \hline

  \end{tabularx}

\end{sidewaystable}

\begin{sidewaystable}[p]

  \caption[ \Xsecstxtnoun of \DstarpTopipDzToKmpip ] {
    \small Differential production \xsecstxtnoun,
    \dxdy{\ensuremath{\sigma}}{\pT}, in $\mub/(\GeVcNS)$ for prompt
    \mbox{\Dstarp $+$ c.c.} mesons in bins of $(\pT, y)$.
    The first uncertainty is statistical, and the second is the total
    systematic.
   \label{tab:xsec:DstarKP}}

  %% $Id: $

  \begin{tabularx}{\linewidth}{C*{5}{d{2}@{\,$\pm$\,}d{2}@{\,$\pm$\,}d{2}}} \hline

    \pT & \multicolumn{15}{c}{$y$} \\

    (\GeVcNS)  & \multicolumn{3}{c}{$(2.0, 2.5)$} & \multicolumn{3}{c}{$(2.5, 3.0)$} & \multicolumn{3}{c}{$(3.0, 3.5)$} & \multicolumn{3}{c}{$(3.5, 4.0)$} & \multicolumn{3}{c}{$(4.0, 4.5)$}\\ \hline

    $(0, 1)$  & \multicolumn{3}{c}{}   & \multicolumn{3}{c}{}   &   26.17 &    5.17 &    3.25 &   36.67 &    6.02 &    4.53 &   46.60 &   12.77 &    6.88 \\
    $(1, 2)$  & \multicolumn{3}{c}{}   &   62.56 &    8.42 &    7.91 &   49.02 &    3.13 &    5.73 &   39.27 &    3.15 &    4.62 &   32.40 &    4.41 &    4.06 \\
    $(2, 3)$  & \multicolumn{3}{c}{}   &   30.60 &    2.85 &    3.66 &   24.93 &    1.54 &    2.91 &   24.11 &    1.77 &    2.86 &   18.55 &    2.37 &    2.45 \\
    $(3, 4)$  &   15.31 &    3.11 &    2.12 &   17.11 &    1.37 &    2.04 &   13.90 &    0.93 &    1.63 &   10.44 &    0.91 &    1.34 &    5.13 &    1.06 &    0.70 \\
    $(4, 5)$  &    \phantom{14}9.90 &    1.61 &    1.35 &    \phantom{114}6.28 &    0.66 &    0.81 &    \phantom{112}6.20 &    0.57 &    0.74 &    \phantom{11}4.51 &    0.53 &    0.59 &    3.41 &    1.02 &    0.52 \\
    $(5, 6)$  &    3.92 &    0.84 &    0.55 &    3.81 &    0.47 &    0.50 &    3.43 &    0.42 &    0.45 &    1.96 &    0.35 &    0.27 & \multicolumn{3}{c}{}   \\
    $(6, 7)$  &    2.40 &    0.59 &    0.36 &    1.78 &    0.32 &    0.24 &    1.05 &    0.25 &    0.15 &    0.68 &    0.24 &    0.10 & \multicolumn{3}{c}{}   \\
    $(7, 8)$  &    1.74 &    0.58 &    0.30 &    1.10 &    0.31 &    0.17 & \multicolumn{3}{c}{}   & \multicolumn{3}{c}{}   & \multicolumn{3}{c}{}   \\

    \hline

  \end{tabularx}

  \vspace{2ex}

  \caption[ \Xsecstxtnoun of \DspTophipip ] {
    \small Differential production \xsecstxtnoun,
    \dxdy{\ensuremath{\sigma}}{\pT}, in $\mub/(\GeVcNS)$ for prompt
    \mbox{\Dsp $+$ c.c.} mesons in bins of $(\pT, y)$.
    The first uncertainty is statistical, and the second is the total
    systematic.
   \label{tab:xsec:DsPhiPi}}

  %% $Id: $

  \begin{tabularx}{\linewidth}{C*{5}{d{2}@{\,$\pm$\,}d{2}@{\,$\pm$\,}d{2}}} \hline

    \pT & \multicolumn{15}{c}{$y$} \\

    (\GeVcNS)  & \multicolumn{3}{c}{$(2.0, 2.5)$} & \multicolumn{3}{c}{$(2.5, 3.0)$} & \multicolumn{3}{c}{$(3.0, 3.5)$} & \multicolumn{3}{c}{$(3.5, 4.0)$} & \multicolumn{3}{c}{$(4.0, 4.5)$}\\ \hline

    $(0, 1)$  & \multicolumn{3}{c}{}   & \multicolumn{3}{c}{}   &   11.23 &    3.64 &    2.48 & \multicolumn{3}{c}{}   & \multicolumn{3}{c}{}   \\
    $(1, 2)$  &   22.50 &    7.79 &    6.09 &   20.41 &    3.07 &    3.53 &   12.04 &    2.10 &    2.36 &   11.00 &    3.09 &    2.61 & \multicolumn{3}{c}{}   \\
    $(2, 3)$  &    \phantom{14}6.03 &    1.88 &    1.43 &    \phantom{114}8.34 &    1.17 &    1.17 &   \phantom{11}10.37 &    1.18 &    1.46 &    \phantom{111}7.34 &    1.31 &    1.22 &    5.89 &    2.22 &    1.42 \\
    $(3, 4)$  &    3.38 &    0.92 &    0.66 &    5.57 &    0.73 &    0.81 &    4.78 &    0.69 &    0.79 &    3.83 &    0.68 &    0.65 &    2.08 &    0.90 &    0.49 \\
    $(4, 5)$  &    1.79 &    0.50 &    0.31 &    2.18 &    0.37 &    0.30 &    1.49 &    0.29 &    0.21 &    1.62 &    0.39 &    0.26 & \multicolumn{3}{c}{}   \\
    $(5, 6)$  &    0.91 &    0.34 &    0.20 &    1.11 &    0.24 &    0.17 &    0.88 &    0.21 &    0.13 &    0.67 &    0.21 &    0.13 & \multicolumn{3}{c}{}   \\
    $(6, 7)$  &    0.68 &    0.23 &    0.15 &    0.51 &    0.16 &    0.08 &    0.62 &    0.18 &    0.10 & \multicolumn{3}{c}{}   & \multicolumn{3}{c}{}   \\
    $(7, 8)$  &    0.60 &    0.21 &    0.14 & \multicolumn{3}{c}{}   &    0.23 &    0.10 &    0.04 & \multicolumn{3}{c}{}   & \multicolumn{3}{c}{}   \\

    \hline

  \end{tabularx}

\end{sidewaystable}

%%% Local Variables: 
%%% mode: latex
%%% TeX-master: t
%%% End: 
         % \label{sec:app:xsec}

\clearpage
\addcontentsline{toc}{section}{References}
\bibliographystyle{LHCb}
\bibliography{main,charm}

\end{document}